\documentclass[sn-aps,iicol]{sn-jnl}% Default with double column layout

\usepackage{amsmath}
\usepackage{bm}% bold math
\usepackage{textcase}

\jyear{2022}%

\theoremstyle{thmstyleone}%
\newtheorem{theorem}{Theorem}%  meant for continuous numbers
\newtheorem{proposition}[theorem]{Proposition}% 

\theoremstyle{thmstyletwo}%
\newtheorem{example}{Example}%
\newtheorem{remark}{Remark}%

\theoremstyle{thmstylethree}%
\newtheorem{definition}{Definition}%

\begin{document}

\title[Article Title]{Uncovering diffusive states of the yeast membrane protein, Pma1, and how  labeling method can change diffusive behavior}


\author[1,2]{\fnm{Mary Lou P.} \sur{Bailey}}
\equalcont{These authors contributed equally to this work.}

\author[1,3]{\fnm{Susan E.} \sur{Pratt}}
\equalcont{These authors contributed equally to this work.}

\author[4]{\fnm{Yongdeng} \sur{Zhang}}

\author[2,3,4,5]{\fnm{Joerg} \sur{Bewersdorf}}

\author[6]{\fnm{Lynne J.} \sur{Regan}}

\author*[1,2,3]{\fnm{Simon G. J.} \sur{Mochrie}}\email{simon.mochrie@yale.edu}

\affil*[1]{\orgdiv{Integrated Graduate Program in Physical and Engineering Biology}, \orgname{Yale University}, \orgaddress{\street{Street}, \city{New Haven}, \postcode{06511}, \state{Connecticut}, \country{USA}}}

\affil[2]{\orgdiv{Department of Applied Physics}, \orgname{Yale University}, \orgaddress{\street{Street}, \city{New Haven}, \postcode{06511}, \state{Connecticut}, \country{USA}}}

\affil[3]{\orgdiv{Department of Physics}, \orgname{Yale University}, \orgaddress{\street{Street}, \city{New Haven}, \postcode{06511}, \state{Connecticut}, \country{USA}}}

\affil[4]{\orgdiv{Department of Cell Biology}, \orgname{Yale University}, \orgaddress{\street{Street}, \city{New Haven}, \postcode{06511}, \state{Connecticut}, \country{USA}}}

\affil[5]{\orgdiv{Department of Biomedical Engineering}, \orgname{Yale University}, \orgaddress{\street{Street}, \city{New Haven}, \postcode{06511}, \state{Connecticut}, \country{USA}}}

\affil[6]{\orgdiv{Institute of Quantitative Biology, Biochemistry and Biotechnology, Center for Synthetic and Systems Biology, School of Biological Sciences}, \orgname{University of Edinburgh}, \orgaddress{\street{Street}, \city{Edinburgh}, \postcode{06511}, \country{UK}}}


\abstract{We present and analyze video-microscopy-based single-particle-tracking measurements of the budding yeast (\textit{Saccharomyces cerevisiae}) membrane protein, Pma1,  fluorescently-labeled either by direct fusion to the switchable fluorescent protein, mEos3.2, or by a novel, light-touch, labeling scheme, in which a 5 amino acid tag is directly fused to the C-terminus of Pma1, which then binds mEos3.2. The diffusivity distributions of these two populations of single particle tracks differ significantly, demonstrating that labeling method can be an important determinant of diffusive behavior. We also applied perturbation expectation maximization (pEMv2) [Physical Review E 94, 052412 (2016)], which sorts trajectories into the statistically-optimum number of diffusive states. For both TRAP-labeled Pma1 and Pma1-mEos3.2, pEMv2 sorts the tracks into two diffusive states: an essentially immobile state  and a more mobile state. However, the mobile fraction of Pma1-mEos3.2 tracks is much smaller ($\sim 0.1$) than the mobile fraction of TRAP-labeled Pma1 tracks ($\sim 0.5$). In addition, the diffusivity of Pma1-mEos3.2's mobile state is several times smaller than the  diffusivity of TRAP-labeled Pma1's mobile state. To critically assess pEMv2's performance,  we compare the diffusivity and covariance distributions of the experimental pEMv2-sorted populations to corresponding theoretical distributions, assuming that Pma1 displacements realize a Gaussian random process. The experiment-theory comparisons for both the TRAP-labeled Pma1 and Pma1-mEos3.2 reveal good agreement, bolstering the pEMv2 approach.}




\maketitle

\section{Introduction}\label{sec1}

%The goals of this paper are two-fold.
The overarching goal of this work is to carefully examine the extent to which the measured
diffusive behavior of a protein of interest (POI) in a heterogeneous,
biological environment can be convincingly described in terms of a limited number of discrete diffusive states, each with its own diffusive properties.
Such diffusive states might correspond to the POI being bound to different binding partners or being located in different local environments. 
To determine the number of diffusive states, we employ the perturbation expectation maximization (pEMv2) software, an unsupervised, systems-level, machine-learning-based data analysis and classification method that sorts a heterogeneous population of single particle tracks. We then developed a method to validate the number of states found by pEMv2, based on the theoretical displacement covariance distributions.
In this paper, we studied the budding yeast ({\em Saccharomyces cerevisiae})
membrane protein, Pma1  \cite{Albers1967,Post1972,Goormaghtigh1986,Morsomme2000,Malinska2003,Kuehlbrandt2004,Spira2012,Mason2013,Henderson2014,Athanasopoulos2019,Heit2021,Zhao2021}
\textemdash in the cell membrane. 
For Pma1, we find that the observed population of single-molecule trajectories
can be well-described in terms of just two diffusive states.
One of these states corresponds to simple diffusion. The other is essentially immobile.

%The goals of this paper are two-fold.
%The first goal is to carefully examine the extent to which the measured
%diffusive behavior of a protein of interest (POI) \textemdash
%in this case, the budding yeast ({\em Saccharomyces cerevisiae})
%membrane protein, Pma1  \cite{Albers1967,Post1972,Goormaghtigh1986,Morsomme2000,Malinska2003,Kuehlbrandt2004,Spira2012,Mason2013,Henderson2014,Athanasopoulos2019,Heit2021,Zhao2021}
%\textemdash in a heterogeneous,
%biological environment \textemdash in this case,  the  cell membrane \textemdash
%can be convincingly described in terms of a limited number of discrete diffusive states, each with its own diffusive properties.
%Such diffusive states might correspond to the protein of interest being bound to different binding partners
%or being located in different local environments. 
%To determine the number of diffusive states in our study, we employ the perturbation expectation maximization (pEMv2) software, an unsupervised, systems-level, machine-learning-based data analysis and classification method that sorts a heterogeneous population of single particle tracks. We then developed a method to validate the number of states found by pEMv2, based on the theoretical displacement covariance distributions.
%For Pma1, we find that the observed population of single-molecule trajectories
%can be well-described in terms of just two diffusive states.
%One of these states corresponds to simple diffusion. The other is essentially immobile.

Spurred by the resulting two diffusive states we find, our second goal is to quantify the differences between the diffusivities and displacement covariances exhibited by Pma1, labeled in two different ways.
%Our second goal is to demonstrate and quantify important differences between
% the distribution of diffusivities and displacement covariances
%exhibited by Pma1, labeled in two different ways,
%even though no obvious differences in the behavior of the two versions of Pma1 are apparent from
%static fluorescence microscopy images.
The first labeling method \textemdash which we refer to as ``TRAP labeling'' \textemdash is a light-touch
method
in which wild-type Pma1 is replaced with a version of Pma1, in which a 5 amino acid tag is directly fused to the C-terminus of Pma1
\cite{Pratt2016,Oi2020,Hinrichsen2017,Oi2020Trap-Peptide,OiProteinScience2020}.
%augmented with a minimally-perturbing
%five-amino-acid C-terminal tag,
This is done in cells expressing a version of a fluorescent protein engineered to bind the tag, thus labeling Pma1.
Second is the commonly used direct-labeling method, which replaces wild-type Pma1 with a Pma1-fluorescent protein (FP) direct-fusion.
In many cases, the assumption that the POI's intrinsic biological function will be unaffected by the direct addition of the FP is surely correct. However,  in some cases, adding an FP to a POI can cause the modified protein to mislocalize \cite{Huh2003} or  misfunction  \cite{Coutu2013, Li2017, Toseland2013, Jensen2012}. For example, it was recently found that tagging dynamin-related protein 1 with GFP alters oligomerization dynamics, causing impaired oligomerization compared to native protein \cite{MontecinosFranjola2020}. In another study, overexpressed, membrane-targeted GFP fusion proteins were found to form organelle aggregates, therefore changing the motion of the protein being studied \cite{Lisenbee2003}. Similarly, Pma1 provides an example of a protein that is sensitive to labeling method: Yeast strains expressing direct fusions of Pma1 and the FPs, mCherry or EGFP, both exhibit compromised growth and mislocalization of Pma1-mCherry or Pma1-EGFP, respectively, to the vacuole~\cite{Hinrichsen2017}. The possibility of mislocalization or misfunction of direct fusion proteins has motivated efforts to develop and deploy alternative
{\em in vivo} labeling methods that are less disruptive of the protein's localization or function \cite{Pratt2016,Oi2020,Hinrichsen2017,Oi2020Trap-Peptide,OiProteinScience2020}.
The different diffusive behavior of these two differently labeled proteins, (between TRAP-labeled and direct fusion proteins) that we observe,
demonstrates that it may be necessary to broadly employ
minimally-perturbing labeling schemes in order to fully realize
and study intrinsic biological behavior in live cells.

This paper is organized as follows. 
In Sec.~\ref{background}, we present necessary background material.
Sec.~\ref{pma1} briefly summarizes what is known about Pma1.
%Sec.~\ref{TRAPlabeling} describes the TRAP labeling method.
In Sec.~\ref{pem}, we review the perturbation Expectation Maximization (pEMv2)  algorithm,
which  determines the number of unique diffusive states within a population of single particle tracks and
sorts individual tracks into those  states. 
In Sec.~\ref{Sec:Theory}, we review theoretical results for probability distributions
of two-dimensional displacement covariance matrix elements~\cite{Bailey2021},
which can be compared to the corresponding experimental distributions.

In Sec.~\ref{SamplePrep},
we describe the  yeast strains employed in this study and how samples were prepared.
We describe the microscopy setup used to collect the data in Sec.~\ref{MicroscopySetup},
and
the methods we used for single particle tracking in Sec.~\ref{SPTMethods}.
%In Sec.~\ref{SimulationMethods}, we describe how we generate simulated particle tracks. 

Sec.~\ref{results} presents and discusses our results.
In Sec.~\ref{microscopyresults}, we  present microscopy images of the strains studied.
In contrast to cells where Pma1 is expressed as a fusion protein with the fluorescent proteins, mCherry and EGFP, 
%Pma1-mCherry and Pma1-EGFP, 
which show defective growth and behavior,
we find that cells expressing Pma1-mEos3.2
do not show a growth defect, nor does Pma1-mEos3.2 mislocalize to the vacuole.
However, these gross observations
do not rule out more subtle differences in behavior between Pma1-mEos3.2 and
TRAP-labeled Pma1.

In Sec.~\ref{Sec:tracks}, we present the experimental track length distributions of TRAP-labeled Pma1
and Pma1-mEos3.2.
In both cases, we find that the track lifetime distribution decreases approximately single-exponentially in time, indicating that Pma1 is a monomer in the cell membrane,
and show that the lifetime of TRAP-labeled Pma1 tracks is shorter than that of Pma1-mEos3.2.
We interpret this difference in lifetime in terms of a non-zero
TRAP-peptide unbinding rate, which allows us to estimate this rate
to be about $6~$s$^{-1}$.

To examine the differences between the diffusive behavior of
TRAP-labeled Pma1 and Pma1-mEos3.2,
we subjected both of these populations of trajectories to pEMv2 analysis
\cite{Koo2015,Koo2016,Rey-Suarez2020}.
In Sec.~\ref{Sec:pEM},
we describe the application of pEMv2 to the population of TRAP-labeled Pma1 single particle trajectories.
pEMv2 sorts TRAP-labeled Pma1 trajectories
%and Pma1-mEos3.2 trajectories
into two states. 
After analyzing the mean squared displacements of the sorted tracks in Sec.~\ref{Sec:MSD}, we test the
sorted populations by comparing their covariance and diffusivity distributions to the corresponding theoretical  distributions.
We  find good agreement between pEMv2-sorted covariance distributions and theory (Sec.~\ref{Sec:experimental_ftirf}),
%providing support for pEMv2's conclusions.
bolstering the pEMv2-based  approach.
%Similar analysis of a simulated population of tracks, that mimics
%experimental TRAP-labeled population (Sec.~\ref{sec:simulated data}) yields excellent agreement between theory and pEMv2-sorted covariance values.
Next, in Sec.~\ref{sec:direct fusion}, we analyze the population of Pma1-mEos3.2 direct fusion tracks 
with pEMv2, and find that it also shows two states.
However,
in contrast to TRAP-labeled Pma1, where the mobile fraction is about 0.5,
the fraction of Pma1-mEos3.2 direct-fusion tracks in the mobile state that is 0.1.
%Specifically, while
%for both labeling methods,
%pEMv2 reveals
%an essentially immobile sub-population, this immobile subpopulation constitutes about 0.5 of the tracks 
%for TRAP-labeled Pma1,  but about 0.9 of the tracks for Pma1-mEos3.2.
%The different immobile fractions
%for TRAP-labeled Pma1 and Pma1-mEos3.2 is part of the difference between the overall diffusivity distributions.
In addition to the much reduced mobile fraction in Pma1-mEos3.2 compared to TRAP-labeled Pma1,
the diffusivities of the mobile sub-populations alone appear to be different for the two labeling strategies with the mean diffusivity of the mobile sub-population of TRAP-labeled Pma1($D_2 \simeq 0.16~\mu$m$^2$s$^{-1}$) 
being several-fold larger than the mean diffusivity   of the mobile state (state 2) of  Pma1-mEos3.2
($D_2 \simeq 0.05~\mu$m$^2$s$^{-1}$).
Direct comparison between the sorted experimental diffusivity and covariance distributions
with the theoretical diffusivity and covariance distributions from
Refs.~\cite{Vestergaard2014, Bailey2021}
reveals good agreement between experiment and theory in this case too.
%
%
%Checking our theory on these results again conveys that pEMv2 has accurately sorted the tracks. 
Finally, in Sec.~\ref{conclusion}, we summarize and conclude.  
%In Appendix A we give more background on the TRAP-labeling method.

\section{Background}
\label{background}
\subsection{Pma1}
\label{pma1}
Pma1 is the most abundant protein in the plasma membrane of budding yeast ({\em Saccharomyces cerevisiae}). It is a transmembrane protein which pumps protons out of the cell and thus plays a role in regulating the pH of the cytoplasm.
Pma1 is also a marker of cell aging because, interestingly, there is less Pma1 in the plasma membranes of newly-budded daughter cells, than in the membranes of their mother cell \cite{Henderson2014}. The yeast plasma membrane is laterally organized into several different membrane ``compartments'' or domains. As the name implies, Pma1 is the majority protein component of the membrane compartment of Pma1 (MCP). Membrane compartments, including MCP, show characteristic linear dimensions of about 0.1~$\mu$m  \cite{Malinska2003,Spira2012}, and differ from each other in their composition, size,  shape, etc. \cite{Athanasopoulos2019} -- MCPs are enriched in sphingolipids, as well as Pma1, and show non-compact, ``network-like'' shapes \cite{Spira2012}. Pma1 contains 918 amino acids, comprising four domains:  a membrane domain which includes ten transmembrane $\alpha$-helices, a phosphorylation domain, a nucleotide-binding domain, and an actuator-domain, which experiences significant rearrangements when Pma1 cycles between the two allosteric states,  activated and inhibited,  involved in its enzymatic  cycle.
The change from Pma1's inhibited to activated state has been proposed to be a consequence of phosphorylation of a specific Ser residue (Ser 899) and the tandem phosphorylation of a Ser/Thr pair (Ser911 and Thr912) \cite{Albers1967,Post1972}. Recently, two cryoelectron microscopy studies of detergent-extracted, lipid-reconstituted, hexamerically-associated Pma1 (from {\em S. cerevisiae} and {\em Neurospora crassa}) provided microscopic details of the inhibited and activated molecular structures \cite{Zhao2021,Heit2021}. Other studies involving {\em in vitro} reconstitution into liposomes \cite{Goormaghtigh1986} or nanodiscs \cite{Justesen2013} report that Pma1 monomers are active in proton pumping. As far as we are aware, no study has yet definitively identified the {\em in vivo} association state of Pma1.

\subsection{Perturbation expectation maximization}
\label{pem}
A number of methods have been introduced for addressing particle tracks exhibiting
biological heterogeneity \cite{Koo2015, Koo2016,Saxton1997, Savin2007, Monnier2015, Das2009, Slator2015,Zhao2019, Pinholt2021}.
In particular,
Refs.~\cite{Koo2015,Koo2016} describe
{\em perturbation expectation maximization},
pEMv2,
which simultaneously analyzes a population of particle trajectories and sorts the trajectories into distinct
diffusive states, each with its own diffusion properties. pEMv2 is a machine-learning approach that makes no {\em a priori} assumptions concerning the character of a diffusive state\textemdash {\em e.g.} whether it corresponds to simple diffusion or not\textemdash but
rather determines each state's diffusive properties directly from the sorted tracks.
%, corresponding to the state in question.

Although pEMv2 performs well on simulations, 
to date
it has been applied to relatively few experimental data sets  \cite{Koo2016, Rey-Suarez2020}.
To further explore pEMv2's performance, we apply it to our experimental dataset of Pma1 single particle tracks. In particular, we compare the statistical properties of pEMv2-sorted populations, which pEMv2 asserts are homogeneous, to theoretical expectations for a population of tracks with a single set of diffusion parameters ~\cite{Vestergaard2014, Bailey2021}.

%Moreover,  in pEMv2's published experimental applications,
%it has found populations consisting of six or more diffusive states.
%To more robustly demonstrate the utility of pEMv2, %  (and similar methods),
%and promote confidence in its results,
%it would be valuable to find an
%experimental application
%that involves just a few diffusive states, and  explore pEMv2's performance in this context,
%in particular,
%by comparing the statistical properties of pEMv2-sorted populations,
%which pEMv2 asserts are homogeneous, to theoretical expectations for a population of tracks
%with a single set of diffusion parameters ~\cite{Vestergaard2014, Bailey2021}.

pEMv2 is described in detail in Refs.~\cite{Koo2015,Koo2016}.
In brief, 
it is an unsupervised,
systems-level, machine-learning-based data analysis and classification method,
that takes as input a heterogeneous population of single particle trajectories.
It
hypothesizes the existence of several different diffusive states within the
population;
it then determines the most likely diffusive properties of each diffusive state, while
sorting each trajectory into the most likely of these diffusive states;
it follows this procedure for different numbers of diffusive states, and
finally picks the optimum number of diffusive states.

Specifically,
for $K$ diffusive states and $M$ tracks, pEMv2 maximizes the log-likelihood of
obtaining the measured tracks:
\begin{equation}
	\log \mathcal{L}
	= \sum_{{m}=1}^{{M}}{\log}\Bigg( \sum_{{k}=1}^{{K}}\pi_k P ( \Delta \mathbf{x}_{m}, \Delta \mathbf{y}_{m} \rvert \mathbf{\Sigma}_k)  \Bigg),
	\label{PopulationLogLikelihood}
\end{equation} 
%\begin{equation}
%	\log \mathcal{L}
%	= \sum_{{m}=1}^{{M}}{\log}\Bigg( \sum_{{k}=1}^{{K}}\pi_k P(\Delta {\mathbf x}_{m}, \Delta {\mathbf y}_{m} | {\mathbf \Sigma}_k)
%	\Bigg),
%	\label{PopulationLogLikelihood}
%\end{equation} 
where
$\Delta {\mathbf x}_m$ is the vector of displacements along $x$ for track $m$,
$\Delta {\mathbf y}_m$ is the vector of displacements along $y$ for track $m$,
$\pi_{k}$ is the fraction of tracks in diffusive state $k$,
${\mathbf \Sigma}_k$ is the displacement covariance matrix of diffusive state $k$ (assumed the same
for $x$ and $y$),
and
\begin{equation}
	P(\Delta {\mathbf x}_m, \Delta {\mathbf y}_m \rvert {\mathbf  \Sigma}_k ) =
	\frac{
		e^{ -\frac{1}{2}\Delta {\mathbf x}_m^T {\mathbf \Sigma}_k^{-1} \Delta {\mathbf x}_m
			-\frac{1}{2}\Delta {\mathbf y}_m^T {\mathbf \Sigma}_k^{-1} \Delta {\mathbf y}_m
		}
	}
	{
		(2\pi)^N \rvert {\mathbf \Sigma}_k \rvert
	}
	\label{IndividualLiklihood}
\end{equation}
is the probability of realizing trajectory $m$,
given that trajectory $m$ corresponds to diffusive state $k$,
with $\rvert{\mathbf \Sigma}_k \rvert$ and ${\mathbf \Sigma}_k^{-1}$  the determinant and inverse of
${\mathbf \Sigma}_k$, respectively,
For trajectories comprising $N$ displacements in $x$ and $y$,
each of $\Delta {\mathbf x}_m$ and  $\Delta {\mathbf y}_m$ is an $N$-component vector
and ${\mathbf \Sigma}_k$ is an $N \times N$ symmetric Toeplitz matrix.
That is, pEMv2 assumes that particle displacements are multivariate Gaussian random variables.
To apply pEMv2 to experimental tracks, we subdivide longer tracks into $N$ step tracks.
pEMv2 maximizes $\log\mathcal{L}$ iteratively by appropriately picking
$\pi_k$ and the matrix elements of ${\mathbf \Sigma}_k$ for each diffusive state $k$, and by assigning each track to  the most likely diffusive state.

Model selection in pEMv2\textemdash that is, picking the appropriate value of $K$\textemdash is implemented by picking the
state with the largest Bayesian Information Criterion (BIC), defined here as
\begin{equation}
	{\rm BIC} = \log \mathcal L -\frac{1}{2} N_P  \log N_D,
	\label{BIC}
\end{equation}
where
$N_D=2NM$ is the number of data points, and
$N_P=KN + K-1$ is the number of model parameters,
equal to the sum of the number of independent covariance matrix elements, $KN$, plus the number of independent population fractions, $K-1$.
The log-likelihood always increases as the number of parameters, and therefore the number of states, increases.
Counteracting this behavior, the second term on the right hand side of Eq.~\ref{BIC} penalizes a larger number of parameters, and therefore a larger number of states. Together these two contributions lead to an optimum value of $K$.

%To analyze the diffusion of labelled
%Pma1 within the cell membrane, we applied pEMv2
%to the population of single particle trajectories obtained for TRAP-labelled Pma1 and Pma1-mEos3.2.

\subsection{Covariance and diffusivity distributions}
\label{Sec:Theory}
According to pEMv2, its sorted
populations each correspond to a single diffusive state with well-defined diffusive properties and parameters.
Assuming that a given diffusive state's displacements
are a zero-mean Gaussian random variable (Eq.~\ref{IndividualLiklihood}),
all statistical properties of the displacements are
determined solely by the mean covariance matrix.
Therefore, to test pEMv2's performance,  we sought to compare the 
pEMv2-sorted covariance- and diffusivity-distributions to the corresponding theoretical 
expectations, given the experimental mean covariances.

The development of theoretical predictions for covariance- and diffusivity-distributions is described in detail in Refs.~\cite{Vestergaard2014,Bailey2021}.
In brief,
for a population of two-dimensional, single-particle trajectories, each of length $N$, and each corresponding to the same diffusive state,
the probability density for a track to yield an estimate of the  covariance matrix element, $n$ steps away from the diagonal, equal to $S_n$, is:
	\begin{multline}
		P(S_n \rvert \mathbf{\Sigma} ) 
		%&= 
		=\int d(\Delta x_1)  ... d(\Delta y_1)... \\
		  P(\Delta { x}, \Delta { y}  \rvert \mathbf{\Sigma} )
		\delta \left (S_n - \frac{1}{2}  \Delta { x}^T {\mathbf C}_n \Delta { x} -  \frac{1}{2}  \Delta { y}^T {\mathbf C}_n \Delta { y} \right ) \\
		=
		\int^\infty_{-\infty} \frac{d \omega}{2 \pi }
		\frac{1}{  { \rvert{ I}+  \frac{i}{2} \omega \mathbf{\Sigma} { C}_n \rvert  }}
		e^{i \omega S_n},
	\label{EQ-16}
\end{multline}
where $[{ C}_0]_{j k} = \frac{2}{N} { I}$
and
$[{ C}_n]_{j k} = \frac{1}{(N-n)} \delta_{j~ k\pm n}$  for $n>0$, where $j=1$ through $N$. 
%
%where ${\mathbf C}_0 = ?$ and ${\mathbf C}_n = ??$ for $n>0$. 

Many experimental systems,
including TRAP-labeled Pma1 and Pma1-mEos3.2,
show diffusive behavior consistent with
simple diffusion with experimental errors,
corresponding to a symmetric, tridiagonal covariance matrix,
where the only non-zero mean covariance matrix elements are on the diagonal, namely $\bar{S}_0$, and
one away from the diagonal, namely $\bar{ S}_1$.
Each individual track yields a measurement of $S_1$ and $S_0$, which are
related to a measurement of the diffusivity, $D$, and  the  static localization noise, $\sigma^2$, for that track via \cite{Vestergaard2014}
%For simple 2D diffusion, the diffusivity  is linearly related to $\Sigma_0$ and $\Sigma_1$ via
\begin{equation}
	S_0  = \sum_{j=1}^N ( \Delta x_j^2 + \Delta y_j^2 ) = 4 D  \Delta t -\frac{4}{3} D  \Delta t_E +2 \sigma^2
	\label{EQ-26}
\end{equation}
and
\begin{equation}
	S_1  = \sum_{j=1}^{N-1} ( \Delta x_j \Delta x_{j+1} + \Delta y_j \Delta y_{j+1} ) = \frac{2}{3}  D  \Delta t_E - \sigma^2,
	\label{EQ-27}
\end{equation}
where
$\Delta t$ is the time between camera exposures and $\Delta t_E$ is the exposure time.
%The static localization noise,  $\sigma^2$,  is the error in particle localization that results from counting a limited number of photons.
The terms involving the exposure time, $\Delta t_E$, correspond to motion blur,
because measurement of the particle position is integrated while the shutter is open.
To be clear, $S_0$, $S_1$, $D$, and $\sigma^2$ are random variables. Their respective
means are  $\bar{S}_0$, $\bar{S}_1$, $\bar{D}$, and $\bar{\sigma}^2$.

Solving
Eqs.~\ref{EQ-26} and \ref{EQ-27} for $D$ and
rewriting in terms of ${\mathbf C}_0$,  ${\mathbf C}_1$, $\Delta {\mathbf x}$ and $\Delta {\mathbf y}$,
\cite{Vestergaard2014,Bailey2021}, we find
	\begin{multline}
		D=\frac{S_0}{4\Delta t}+\frac{S_1}{2 \Delta t} \\
		=
		\frac{1}{4 \Delta t} \left ( \Delta { x}^T { C}_0  \Delta { x}
		+
		\Delta { y}^T { C}_0  \Delta { y} \right ) \\
		+ \frac{1}{2 \Delta t} \left (  \Delta { x}^T { C}_1 \Delta { x}
		+  \Delta { y}^T { C}_1 \Delta { y} \right ).
		\label{EQ-28}
	\end{multline}

	Similarly,
	\begin{multline}
		\sigma^2=   
		\frac{\Delta t_E}{ 6 \Delta t} 
		%	\frac{1}{6}
		S_0
		%-\frac{2}{3}
		+ \left ( \frac{\Delta t_E}{3 \Delta t} - 1 \right ) 
		S_1	\\
		=
		\frac{\Delta t_E}{ 6 \Delta t} 
		%\frac{1}{6}
		\left ( \Delta { x}^T { C}_0  \Delta { x} 
		+
		\Delta { y}^T { C}_0  \Delta { y} \right ) \\
		+  \left ( \frac{\Delta t_E}{3 \Delta t} - 1 \right )
		%	-\frac{2}{3}
		\left (  \Delta { x}^T { C}_1 \Delta { x} 
		+  \Delta { y}^T { C}_1 \Delta { y} \right ).
	\end{multline}
	%Because of the physical importance of the diffusivity, we also consider its distribution, mean, variance and third moment for a distribution of D values estimated from tracks using Eq.~\ref{EQ-28}.
It follows that the
the probability densities of the diffusivity and
the static localization noise can be expressed as
\begin{equation}
	\begin{array}{ll}
		P(D \rvert  \mathbf{\Sigma} ) = 
		\int^\infty_{-\infty} \frac{d \omega}{2 \pi }
		\frac{1}{  { \rvert{ I}+  \frac{i}{8 \Delta t} \omega  \mathbf{\Sigma} ({ C}_0+2 { C}_1) \rvert  }}
		e^{i \omega D}.
	\end{array}
	\label{EQ-9}
\end{equation}
and
\begin{equation}
	\begin{array}{ll}
		P(\sigma^2 \rvert  \mathbf{\Sigma} ) = 
		\int^\infty_{-\infty} \frac{d \omega}{2 \pi }
		\frac{1}{  { \rvert{ I}+  \frac{i}{2 \Delta t} \omega  \mathbf{\Sigma} (  \frac{\Delta t_E}{ 6 \Delta t} { C}_0+ \left ( \frac{\Delta t_E}{3 \Delta t} - 1 \right ) { C}_1) \rvert  }}
		e^{i \omega D},
	\end{array}
	\label{EQ-10}
\end{equation}
respectively.
For each of Eq.~\ref{EQ-16}, Eq.~\ref{EQ-9}, and Eq.~\ref{EQ-10},
to provide explicit results,
we calculate the determinant in the integrand  as a function of $\omega$
and then carry out each integral over $\omega$ numerically.

Eq.~\ref{EQ-16}, Eq.~\ref{EQ-9}, and Eq.~\ref{EQ-10} are applicable when there is one diffusive state.
The generalization to $K$ states,
with population fractions specified by $\{\pi_k \}$ and diffusion properties specified by  $\{ {\mathbf \Sigma}_k \}$,
is straightforward:
%\begin{widetext}
\begin{equation}
	\begin{array}{ll}
		P(S_n\rvert \{\pi_k\},\{{\mathbf \Sigma}_k\}) =
		\sum_{k=1}^{{K}} \pi_k  P(S_n \rvert \mathbf{\Sigma}_k ),
		% P(S_{n_1} \rvert \mathbf{\Sigma}_1 ) + P(S_{n_2} \rvert \mathbf{\Sigma}_2 )
		%	&= 
		%	A \int^\infty_{-\infty} \frac{d \omega}{2 \pi }
		%	\frac{1}{  { \rvert{ I}+  \frac{i}{2} \omega \mathbf{\Sigma}_1 { C}_n \rvert  }}
		%	e^{i \omega S_{n1}} +  B	\int^\infty_{-\infty} \frac{d \omega}{2 \pi }
		%	\frac{1}{  { \rvert{ I}+  \frac{i}{2} \omega \mathbf{\Sigma}_2 { C}_n \rvert  }}
		%	e^{i \omega S_{n2}}
	\end{array}
	\label{EQ-11}
\end{equation}
\begin{equation}
	\begin{array}{ll}
		P(D\rvert \{\pi_k\},\{{\mathbf \Sigma}_k\}) =
		\sum_{k=1}^{{K}} \pi_k  P(D \rvert \mathbf{\Sigma}_k ),
		% P(S_{n_1} \rvert \mathbf{\Sigma}_1 ) + P(S_{n_2} \rvert \mathbf{\Sigma}_2 )
		%	&= 
		%	A \int^\infty_{-\infty} \frac{d \omega}{2 \pi }
		%	\frac{1}{  { \rvert{ I}+  \frac{i}{2} \omega \mathbf{\Sigma}_1 { C}_n \rvert  }}
		%	e^{i \omega S_{n1}} +  B	\int^\infty_{-\infty} \frac{d \omega}{2 \pi }
		%	\frac{1}{  { \rvert{ I}+  \frac{i}{2} \omega \mathbf{\Sigma}_2 { C}_n \rvert  }}
		%	e^{i \omega S_{n2}}
	\end{array}
	\label{EQ-12}
\end{equation}
and
\begin{equation}
	\begin{array}{ll}
		P(\sigma^2 \rvert \{\pi_k\},\{{\mathbf \Sigma}_k\}) =
		\sum_{k=1}^{{K}} \pi_k  P(\sigma^2 \rvert \mathbf{\Sigma}_k ).
		% P(S_{n_1} \rvert \mathbf{\Sigma}_1 ) + P(S_{n_2} \rvert \mathbf{\Sigma}_2 )
		%	&= 
		%	A \int^\infty_{-\infty} \frac{d \omega}{2 \pi }
		%	\frac{1}{  { \rvert{ I}+  \frac{i}{2} \omega \mathbf{\Sigma}_1 { C}_n \rvert  }}
		%	e^{i \omega S_{n1}} +  B	\int^\infty_{-\infty} \frac{d \omega}{2 \pi }
		%	\frac{1}{  { \rvert{ I}+  \frac{i}{2} \omega \mathbf{\Sigma}_2 { C}_n |  }}
		%	e^{i \omega S_{n2}}
	\end{array},
	\label{EQ-13}
\end{equation}
where $\{..\}$ indicates ``the set of  ...''.

\section{Materials and Methods}
\label{mandm}
\subsection{Sample preparation}
\label{SamplePrep}
All measurements described in this paper employed strains of the budding yeast, {\em Saccharomyces cerevisiae}.
The construction of these strains
%were completed by our collaborators in the Regan Lab (Yale University prior to summer 2018) and 
is described in detail in  \cite{Hinrichsen2017,Hinrichsen2018}.
For  strains expressing a modified version of Pma1,
we replaced the native chromosomal PMA1 gene with a gene encoding modified Pma1. Modified Pma1 was expressed from the endogenous PMA1 promoter, allowing us to study behavior at endogenous expression levels.

For our microscopy experiments, overnight cultures were grown in synthetic defined (SD) media, with $2\%$ sucrose and $1\%$ raffinose. 
These starter cultures were diluted into fresh media, containing $2\%$ galactose, to obtain a final OD$_{600}\approx0.05$. Growth was then continued at $30^{\circ}$C for a further 8 hrs. Cells from these cultures were then imaged as follows:
1 mg/mL of concanavalin A (conA) was applied to a clean cover slip, and incubated at room temperature for 10 minutes. Then, 0.5-1.0~mL of fresh milli-Q water was used to rinse off the excess conA. Next, the yeast culture, previously vortexed for approximately $30$~s to separate any cell clusters, was added to the conA-coated cover slip and incubated at room temperature for an additional 10 minutes. Excess, unbound cells were rinsed from the cover slip, which was then sealed to a microscope slide using a 1:1:1 ratio mixture of Vaseline, lanolin and paraffin wax (VALAP).

\subsection{Microscopy}
\label{MicroscopySetup}
Microscopy measurements
to track the motions of individual molecules of TRAP-labeled Pma1
and
Pma1-mEos3.2
were carried out using the custom-built microscope,
described in Ref.~\cite{Zhang2017}, which has
both total-internal reflection fluorescence (TIRF) and photo-activation localization
microscopy (PALM) capabilities.
%in the Bewersdorf lab at Yale University as 
We employed a $405$ nm-wavelength laser to switch mEos3.2 into its red state and a $560$ nm-wavelength laser for imaging of switched mEos3.2.
As noted previously, Pma1 is the most abundant yeast membrane protein.
Therefore,
switching a subset of the population into mEos3.2's red fluorescent state ensures sufficiently
isolated, and therefore resolvable, individual trajectories for Pma1-mEos3.2, that
are suitable for unambiguous single particle tracking.
The trajectories reported and analyzed in this paper were collected under
conditions of TIRF illumination, which restricted switching and fluorescence excitation to
the portion of a cell's membrane in close proximity to the coverslip.
In addition, however, a limited quantity of wide-field PALM data 
were collected to visually assess the extent to which PALM signal is associated with the cell membrane.
The intensity of the $405$ nm-wavelength laser was manually adjusted during the experiments
%  from mean intensity values of $0$ to $1$ kW/cm$^{2}$
to  ensure a sufficient signal rate  throughout the duration of data acquisition.
% The $560$ nm-wavelength laser was set to a mean intensity value of either $\sim$$1$ or $2$ kW/cm$^{2}$.
% We used an oil-immersion objective with $100\times$ magnification and numerical aperture of $1.49$,
%allowing us to carry out
%both widefield and 
%TIRF PALM experiments.
The fields-of-view imaged were $256\times256$ pixels, with square pixels,
each spanning $103$~nm on a side.
%A representative single movie frame is shown in Fig.~\ref{ExAnalysisPALM}A).
Images were collected at a rate of $100$ frames per second (fps), corresponding to $\Delta t =0.01$~s.
The exposure time was also $\Delta t_E = 0.01$~s.
%We generally aimed to collect movies consisting of
% $\sim 60000$ frames from each field of view.
A custom, reflection-based autofocusing system
was deployed during data acquisition to maintain the microscope focus.
%We also collected  brightfield images  pre- and post-PALM data acquisition to assess any xy-drift.
% Data were excluded from analysis if too much drift was present.

\begin{table}
	%\begin{center}
	\begin{tabular}{| c | c |c|}
		\hline 
		& TRAP-peptide- &Pma1-  \\ 
		& labeled Pma1 &mEos3.2  \\ \hline
		Number  of & 14  & 20 \\ 
		cells studied &   &  \\ \hline
		Number  &59226  &20253 \\
		of tracks &  &\\ \hline
		Number of & 60000  & 60000 \\ 
		movie frames &   &  \\ \hline
		
	\end{tabular}
	\caption{\label{Table1} Number of cells, number of tracks
		(defined by two or more steps), and number of movie frames for the two strains imaged under TIRF
		conditions.}
\end{table}

\subsection{Single particle tracking}
\label{SPTMethods}
Single particle tracking  was accomplished using a locally-customized  Matlab version (The MathWorks, Inc.)
of the software, described in Ref.~\cite{Crocker1996},
initially resulting in a number of candidate localizations
in each movie frame.
%(Fig.~\ref{ExAnalysisPALM}B).
%Owing to their large size, movies were first split into smaller movies of $500$ frames each.
%Each cropped movie was initially denoised using a wavelet decomposition approach, and then spatially filtered. A first pass of rough peak localization was carried out next, followed by calculations of the centroids of the likely peaks. The centroids were used in conjunction with symmetric $2$D Gaussian fitting to obtain the final sub-pixel estimates of the signal localizations (Figure \ref{ExAnalysisPALM}B). We then thresholded the pixel positions by the widths from the Gaussian fitting (set the lower bound to be $\sigma=0.5$ and the upper bound to be $\sigma=3.5$), and determined the single particle tracks (single particle trajectories). The single particle trajectories were found using an algorithm that looks at all possible pixel positions at some time, t\textsubscript{i}, and then minimizes the total squared displacement between those pixel positions and all possible pixel positions that happen some time t\textsubscript{i$+$1} later.
To achieve our final set of localizations,
we manually segmented brightfield images to find cells and
excluded any (apparent) localizations outside cells.
We also excluded localizations with
standard deviations less than 0.5 pixels or more than 3.5 times the mean
standard deviation of the population of localizations.
Finally, to construct single-particle trajectories, spanning multiple movie frames\textemdash {\em i.e.} spanning time\textemdash
we insisted that the maximum number of pixels 
that a particle can move between successive frames is 2 pixels,
and that the maximum number of frames for which a particle can transiently disappear and still be considered part of a specific trajectory is 1 frame.

\begin{figure}[t!]
	\includegraphics[scale=1]{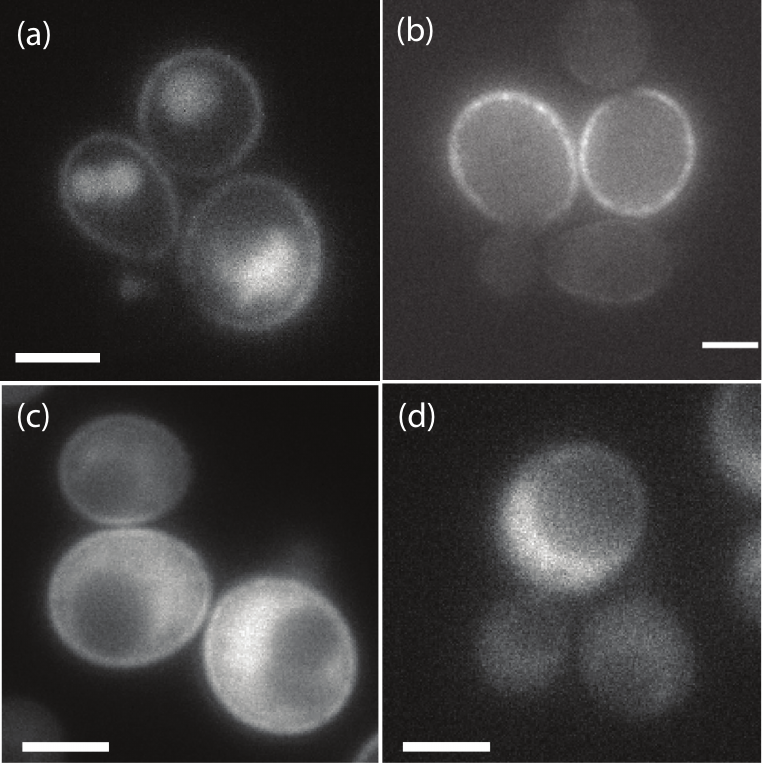}% Here is how to import EPS art
	\caption{\label{}Wide-field fluorescence microscopy images for four strains of
		{\em S. cerevisiae}:
		(a) Cells expressing the Pma1-mCherry direct fusion;
		(b) Cells expressing the Pma1-mEos3.2 direct fusion;
		(c) Cells expressing the Pma1-MEEVF/TRAP-mEos3.2 TRAP-peptide system;
		and
		(d) Cells expressing untagged Pma1 and TRAP-mEos3.2.
		% A) is a positive control and B) is a negative control.
		In each case, the scale bars correspond to $3~\mu$m.
		\label{Figure1}  }
\end{figure}

%\subsection{Simulations}
%\label{SimulationMethods}
%Populations of simulated trajectories were generated as described in Ref.~\cite{Koo2016}.
%In brief, first, tracks undergoing normal diffusion were generated
%according to:
%\begin{equation}
%	x(i+1) = x(i) + [2D\Delta t_S\delta_{i,j}]^{\frac{1}{2}}W(j)
%	\label{DiffusionSimulation}
%\end{equation}
%where $x(i)$ is the $x$-coordinate of the particle's position at time step $i$, $x(0)=0$,
%$W(j)$ is a standard Brownian motion with  $\langle W(j) \rangle = 0$ and $\langle W(j)W(j) \rangle = \delta_{i,j}$,
%and $\Delta t_S = \frac{1}{32} \Delta t_E$ is the simulation time step.
%We then averaged over blocks of 32 simulation time steps,  thus achieving a simulated motion-blurred trajectory
%with successive positions, separated in time by the experimental exposure time.
%Finally, we added Gaussian-distributed localization noise to each position in the motion-blurred trajectory.

\begin{figure} 
	\includegraphics[scale=0.5]{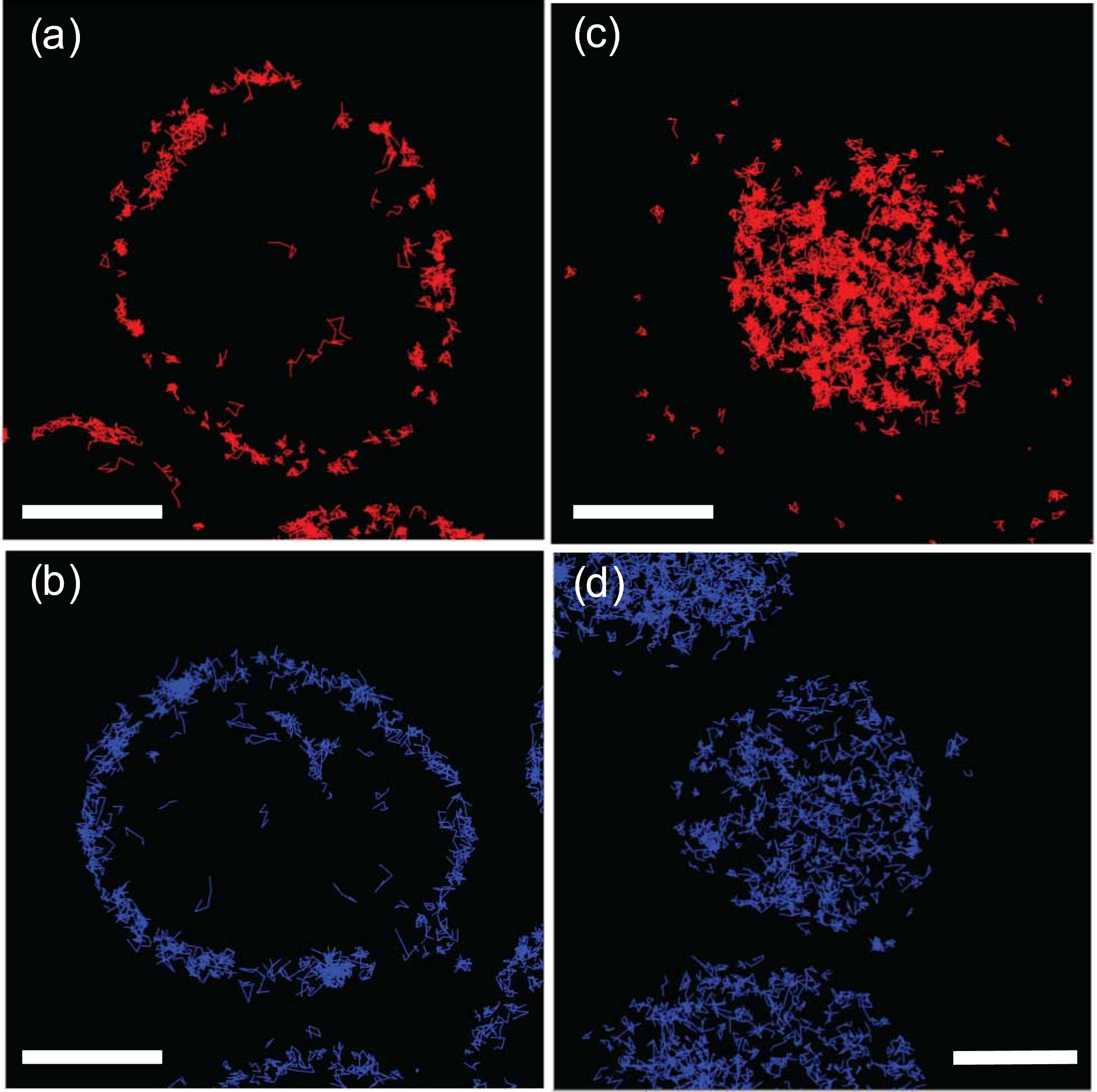}% microscopy_tracks.pdf
	\caption{
		Trajectories comprising four or more steps obtained from 20000 frames of wide-field PALM movies from cells with (a) Pma1-mEos3.2 and (b) TRAP-labeled Pma1. Trajectories comprising four or more steps obtained from 20000 frames of PALM movies,
		under TIRF illumination, from cells with (c) Pma1-mEos3.2 and (d) TRAP-labeled Pma1. In each panel, the scale bar represents 3 $\mu m$.
		\label{Figure2_comb} }
\end{figure}

\section{Results and Discussion}
\label{results}
\subsection{Microscopy}
\label{microscopyresults}
Fig.~\ref{Figure1}(a) shows wide-field fluorescence microscopy images of
cells expressing a Pma1-mCherry direct fusion.
Apparent in this image is a bright halo of fluorescence intensity  at
the cell periphery, corresponding to the localization of Pma1-mCherry to the cell membrane. 
Also apparent in this image is fluorescence intensity originating from regions inside the cell,
corresponding to mislocalization
of Pma1-mCherry to vacuoles.
Evidently, the fusion of mCherry to Pma1 causes Pma1's behavior to diverge significantly from
it intrinsic behavior, which does not involve localization to vacuoles.
Fig.~\ref{Figure1}(b) shows  cells expressing a Pma1-mEos3.2 direct fusion. In this case also, there is
a halo of fluorescence intensity  at
the cells' periphery, corresponding to the localization of Pma1-mEos3.2 to the cell membrane.
However, in this case, there is no mislocalization to vacuoles.
Fig.~\ref{Figure1}(c) shows cells with TRAP-labeled Pma1, {\em i.e.} cells
expressing both Pma1-MEEVF and TRAP-mEos3.2.
In this case too, there is a halo at the cell membrane,
indicating the presence of the Pma1-MEEVF-TRAP-mEos3.2 complex
localized to the membrane, consistent with strong  MEEVF-TRAP binding, for which
Ref.~\cite{Pratt2016} estimated the
dissociation constant to be $K_D \simeq 300$~nM.
In Fig.~\ref{Figure1}(c), fluorescence intensity also originates within the cell,
corresponding to free TRAP-mEos3.2 in solution.
Finally, Fig.~\ref{Figure1}(d) shows cells expressing TRAP-mEos3.2 but with unmodified Pma1.
As expected, there is no longer an intensity halo at the cells' periphery,
because the TRAP-mEos3.2's MEEVF binding partner is absent from Pma1 or anywhere else.

\begin{figure} %BIC N=9 FTIRF
	\includegraphics[scale = 0.455]{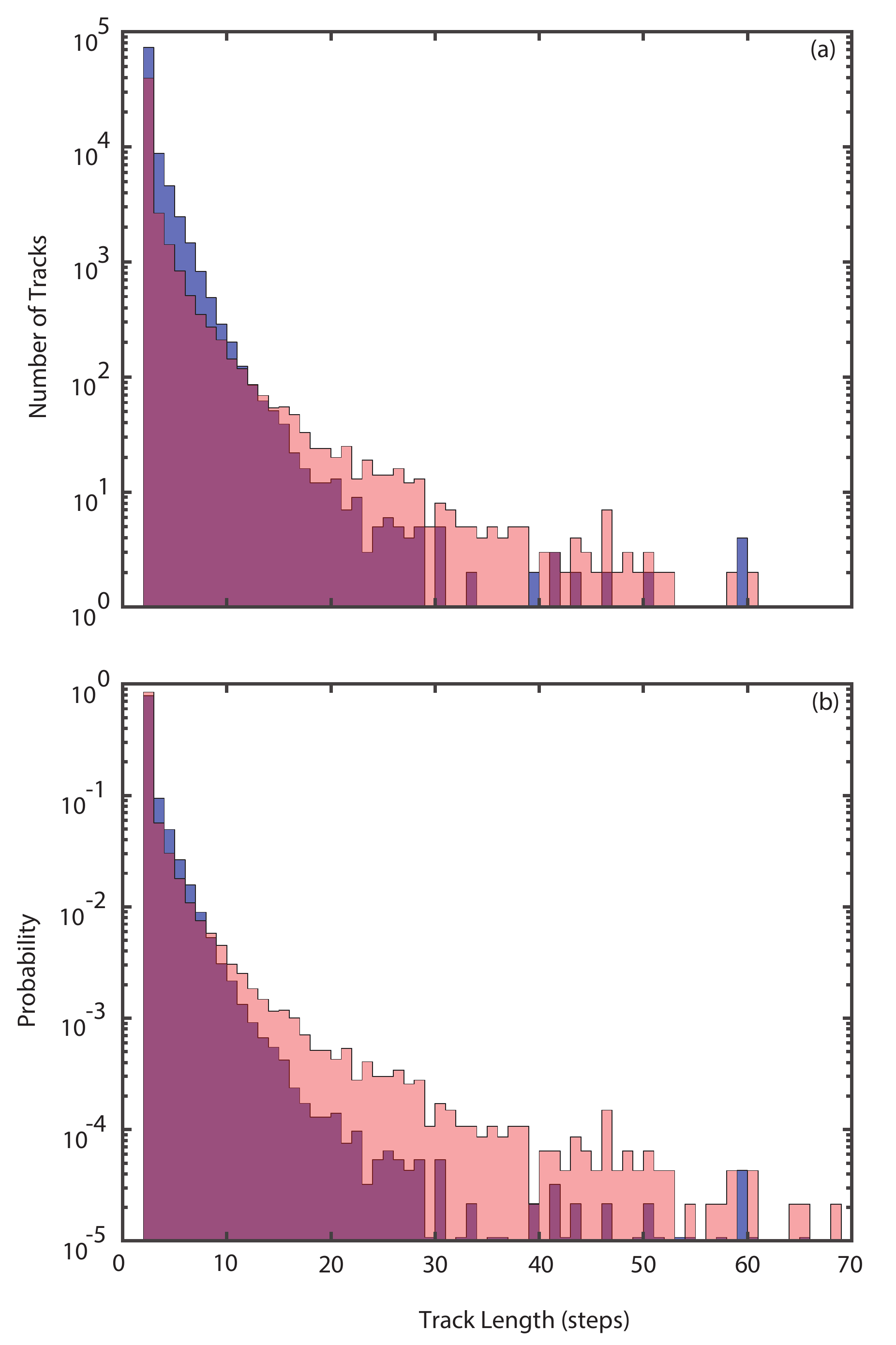}% Here is how to import EPS art
	\caption{\label{fig:tracks} (a) Number of tracks and (b) probability versus track length plotted on logarithmic-linear axes
		for TRAP-labeled Pma1, shown in blue, and Pma1-mEos3.2, shown in pink. 
		Two is the minimum number of connected steps to be considered a track.
	}
\end{figure}

For the images in Fig.~\ref{Figure1}(b-d), mEos3.2 was visualized through its unswitched green fluorescence.
However, mEos3.2
% is photoactivatable and
can be switched by exposure to 405~nm wavelength light into
a red fluorescent state.
Fig.~\ref{Figure2_comb}(a-b) depicts trajectories of  switched, red-emitting proteins containing four or more steps obtained from movies collected under 565-nm-illumination from cells with (a) Pma1-mEos3.2 and (b) TRAP-labeled Pma1.
In both cases, the spatial distribution of trajectories clearly outlines the periphery of the cell that intersects the focal plane.
There is no obvious difference in the spatial distribution of the tracks obtained for
Pma1-mEos3.2, which we expect to be strictly confined to the cell membrane,
and tracks obtained for TRAP-labeled Pma1.
Initially, it may seem surprising that there are no trajectories corresponding to
the cytocellular fluorescence apparent in Fig.~\ref{Figure1}(c), which depicts the same yeast strain
as Fig.~\ref{Figure2_comb}(b).
However, 
we expect that an unbound TRAP-mEos3.2, freely and rapidly diffusing in the
cytosol, would yield either a localization in just a single movie frame or a very short track at most.
We infer, therefore, that trajectories with four or more steps
in cells with TRAP-peptide labeling overwhelmingly must correspond to
labeled Pma1 in the cell membrane, and not to unbound MEEVF-mEos3.2 in the cytosol.

More generally, the observation that 
unbound TRAP-mEos3.2 may be eliminated from consideration by insisting on sufficiently long tracks
suggests a possible general strategy for {\em imaging} TRAP-labeled POIs \cite{Oi2020,Oi2021},
namely to insist that tracks endure for some minimum number of time steps.
In this case, the fluorescence from unbound TRAP-FPs
may be less of a problem for imaging than it might originally have seemed.

Fig.~\ref{Figure2_comb}(c-d) shows  trajectories also containing four or more steps, obtained from PALM movies collected with TIRF illumination, from cells with (c) Pma1-mEos3.2 and (d) TRAP-labeled Pma1.
For both labeling methods, tracks appear largely confined to roughly circular regions.
Because the tracks are derived from TIRF data,
so that we expect to visualize only tracks close to the cover slide,
we identify each roughly-circular region as the
area of the cell in contact with the cover slide.
Apparent tracks outside these regions were eliminated from further consideration.
Because all of these trajectories involve four or more steps and because they were acquired through TIRF,
we are confident that overwhelmingly they correspond to TRAP-labeled Pma1 and not freely-diffusing TRAP-mEos3.2.
Although wide-field PALM measurements were important for establishing the conditions necessary to examine
TRAP-labeled Pma1, in the remainder of the paper,
we focus solely on the two-dimensional trajectories collected with TIRF illumination, in order to
avoid any ambiguities associated with the sideways view of membrane-based trajectories
acquired in wide-field.

\subsection{Distribution of track lengths}
\label{Sec:tracks}
Fig.~\ref{fig:tracks} displays the number of tracks versus track length on linear-logarithmic axes for both TRAP-labeled Pma1
(blue)
and Pma1-mEos3.2 (pink).
Evidently, for both labeling methods, the number of tracks is a monotonically decreasing function of the number of steps in the track.
The track length distribution is equivalent to the distribution of fluorescent state lifetimes,
because the product of the number of steps in a track and duration of a step corresponds to the lifetime of the
corresponding FP(s).
%Although the track-length distribution is not quite single exponential, which would appear linear on linear-logarithmic axes,
The behavior of the lifetime distribution clearly indicates that each of the observed tracks corresponds to a single FP, {\em i.e.}, these are monomer tracks.
In contrast to the observations, the lifetime distribution for a protein complex, which includes several FPs moving together,
is expected to start at zero, then increase to a peak at non-zero time, before decreasing to zero again at large times.
This prediction is readily understood in the  case of a monomer lifetime distribution
given by 
%$\lambda P_1$,
$\lambda e^{-\lambda t}$ --  a  simple exponential with a characteristic rate, $\lambda$.
%where $P_1=e^{-\lambda t}$ is the probability that a monomer does not bleach between time $0$ and time $t$.
Then, for an $n$-mer comprising $n$ independent monomers, the lifetime distribution is readily shown to be 
%$n \lambda P_1 (1-P_1)^{n-1} = 
$n \lambda e^{-\lambda t} (1-e^{-\lambda t})^{n-1} $.
With increasing $t$, this $n$-mer lifetime distribution indeed starts at zero at $t=0$, then increases
to a peak at $t=\frac{\log n}{k}$, before decreasing to zero again at large times, as advertized.

Although the tracks of both TRAP-labeled Pma1 and Pma1-mEos3.2 correspond to monomers,
there is nevertheless a clear difference in the track length distribution between the two methods with the
direct fusion showing more long tracks than TRAP-labeled Pma1.
This observation
is consistent with the interpretation that
when Pma1 is directly bound to the FP, the lifetime of the signal is limited by photobleaching, whereas
when Pma1 is labeled by the TRAP-peptide pair, the lifetime is determined by a combination of photobleaching and the lifetime of
TRAP-MEEVF binding.
Thus, we expect shorter lifetimes for TRAP-labeled Pma1 than for Pma1-mEos3.2. In fact, the normalized distribution of TRAP-labeled Pma1 track lengths (Fig.~\ref{fig:tracks}(b))
falls a factor of $e$ below that of Pma1-mEos3.2 for a track length of about 15, suggesting that the
mean
lifetime of the MEEVF-TRAP bond is about 0.15~s.
Equivalently, the MEEVF-TRAP unbinding rate is about 6~s$^{-1}$.

\subsection{Population-averaged diffusivity distributions}

\begin{figure} 
	\includegraphics[scale=0.445]{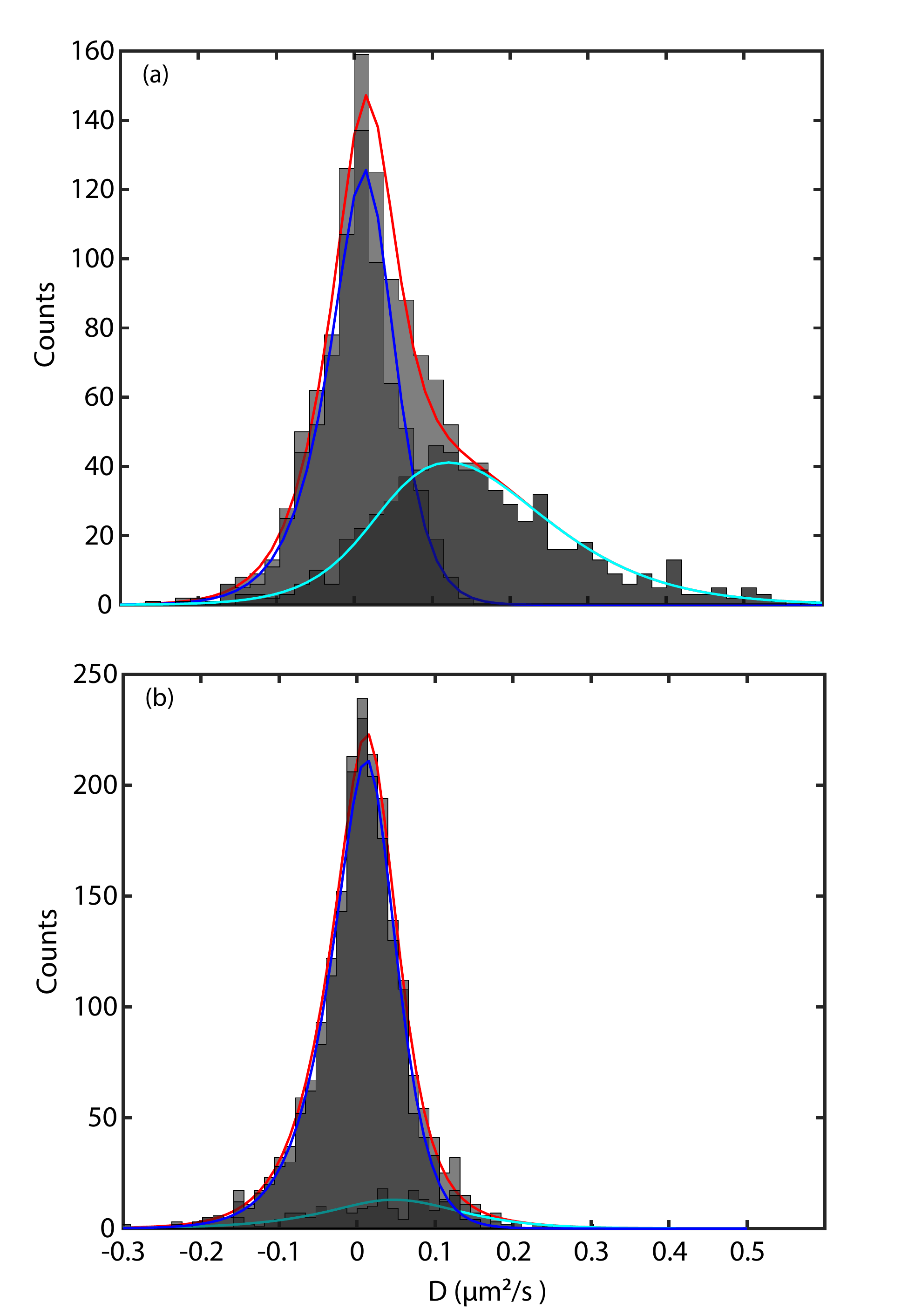} 
	\caption{
		\label{fig:D_exp}
		Comparison between
		the  distributions of diffusivities, $D$,  for (a) experimental TRAP-labeled Pma1
		and (b) experimental direct fusion data.
		In both cases, the track length equals 9 steps. 
		The overall, unsorted, population-averaged distributions, shown as the light grey histograms, are plotted with the theoretical two-component curve, shown as the red curve, calculated using Eq.~\ref{EQ-12} and the pEMv2-found covariance values.
		The sorted diffusivity distributions (dark and darker grey histograms) are shown with their corresponding single state theory curves (blue and cyan curves), given by Eq.~\ref{EQ-9}.
	}
\end{figure}

Fig.~\ref{fig:D_exp} compares population-averaged, diffusivity distributions,
calculated on the basis of Eq.~\ref{EQ-28}, for
(a) the population of
TRAP-labeled Pma1 tracks
and  (b) the population of Pma1-mEos3.2 tracks.
In the figure, the overall diffusivity distributions of these populations are shown as the light-grey histograms.
Although both protein variants' overall diffusivity distributions are peaked near zero,
the Pma1-mEos3.2's diffusivity distribution is largely confined within $\pm 0.1$~$\mu$m$^2$/s.
By contrast,
TRAP-labeled Pma1's diffusivity distribution shows a large diffusivity tail, extending beyond $0.5$~$\mu$m$^2$/s.
Evidently, there is a substantial, qualitative  difference between the overall diffusivity distribution of Pma1-mEos3.2
and the diffusivity distribution of TRAP-labeled Pma1.
This result  shows directly, without any further detailed analysis,
that labeling strategy significantly affects the dynamics of membrane-bound Pma1,
with unknown consequences to Pma1's biological roles.
Because TRAP-labeling minimally modifies the protein of interest, and direct fusions with other fluorescent
proteins causes growth defects and Pma1 mislocalization, our assumption is that TRAP-labeled Pma1
more closely represents the intrinsic biological behavior of wild-type Pma1, than does Pma1-mEos3.2.

\subsection{pEMv2 sorts TRAP-labeled Pma1 trajectories into two diffusive states}\
\label{Sec:pEM}
pEMv2 is most reliable for a large number of long trajectories \cite{Koo2016}.
By contrast, experimentally, as illustrated in Fig.~\ref{fig:tracks},
the number of available tracks decreases rapidly with increasing track length.
Therefore, to explore pEMv2's performance and consistency
over the range of available track lengths and numbers of tracks,
we chose to  partition each dataset into populations of trajectories with track lengths
of 5, 6, 7, 8, 9, 10, 11, 12, and 15 steps.
For the entire population of tracks, we find that the mean covariances, $\bar{S}_n$, are very close to zero for all $n>1$ for all track lengths.
Therefore, for simplicity we chose to set these covariances identically equal to zero when running pEMv2.
The statement that $\bar{S}_n=0$ for $n>1$ is equivalent to the statement that
labeled Pma1 undergoes simple diffusion in the presence of experimental errors.
%As shown below, this assumption is justified {\em a postiori} by the observation of
%mean-square-displacements that increase linearly with delay time, which is behavior consistent
%with simple diffusion with experimental errors.

\begin{figure} %BIC N=9 FTIRF
	\includegraphics[scale=0.455]{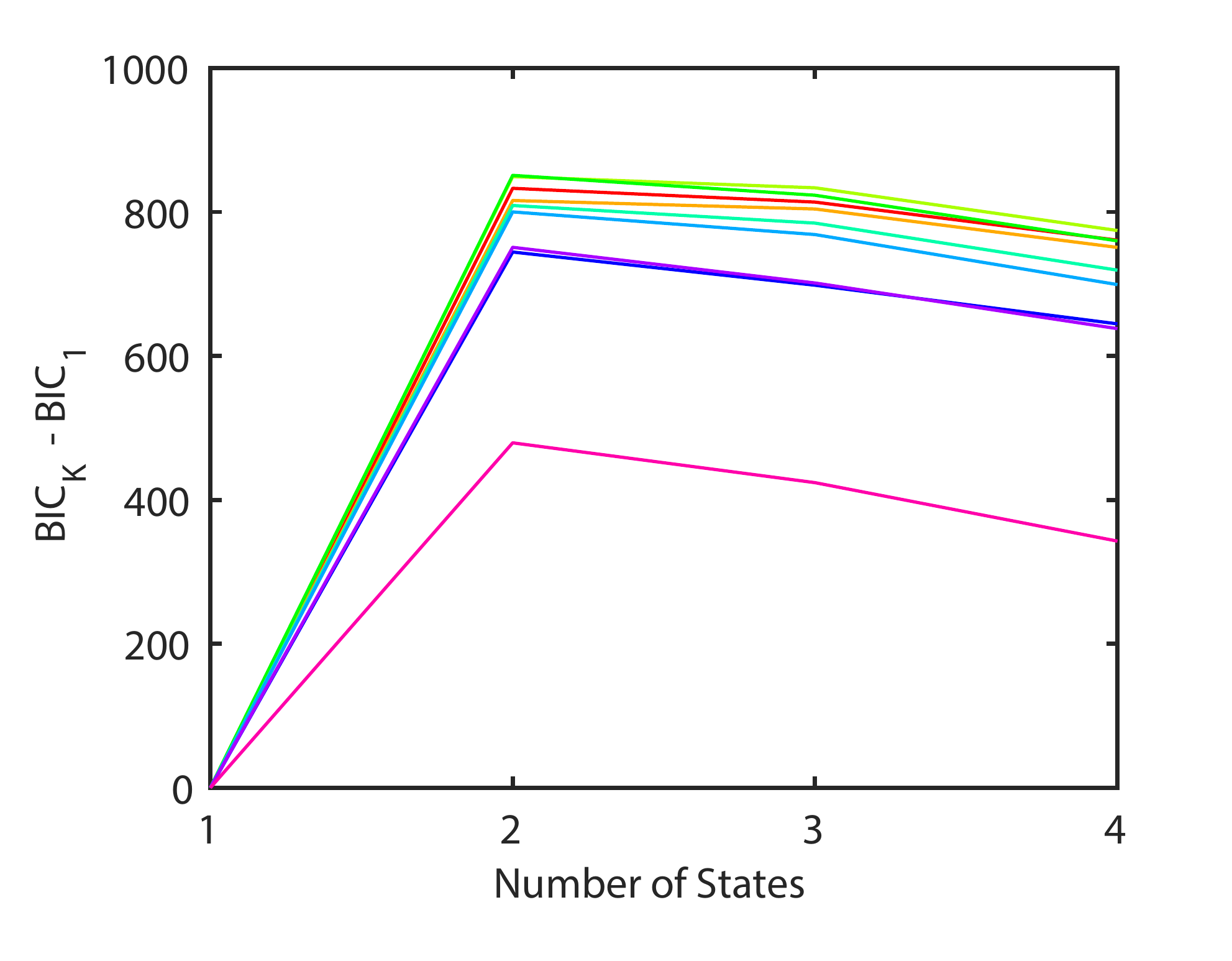}% Here is how to import EPS art
	\caption{\label{fig:ftirf_bic}
		Relative Bayesian Information Criterion (BIC$_K$-BIC$_1$) values
		versus number of diffusive states for TRAP-labeled Pma1 trajectories of
		length 5 (red), 6 (orange), 7 (lime green), 8 (green), 9, (blue-green), 10 (blue), 11 (dark blue), 12 (purple), and 15 (magenta).
		\color{black} 
		The highest relative BIC value for each track length occurs for two states, irrespective of track length. }
\end{figure}

We  carried out a pEMv2-analysis for each population of different-length trajectories.
Fig.~\ref{fig:ftirf_bic} plots the values of BIC$_K-$BIC$_1$ returned by pEMv2  versus the number of diffusive states, $K$,
for track lengths of 5, 6, 7, 8, 9, 10, 11, 12 and 15 steps.
For each track length, the BIC achieves a maximum for $K=2$, irrespective of track length,
indicating  that pEMv2 robustly supports  two subpopulations, namely state 1 and state 2 -- with distinct diffusive properties.

\begin{table*}
	%\begin{center}
	\centering\resizebox{\textwidth}{!}
	{\begin{tabular}{| c | c | c | c | c | c | c | c | c | c | c | c | c | c | c | c | c | c | c |}
			\hline
			{Track Length} & ${\bar{S}_{01}}$ & ${\bar{S}_{01_{err}}}$ & ${\bar{S}_{11}}$ & ${\bar{S}_{11_{err}}}$ & $\bar{D_1}$ & ${D_{1_{err}}}$ & ${\sigma^2_1}$ & ${\sigma^2_{1_{err}}}$ & ${\bar{S}_{02}}$  & ${\bar{S}_{02_{err}}}$ & ${\bar{S}_{12}}$  & ${\bar{S}_{12_{err}}}$ & $\bar{D_2}$ & ${D_{2_{err}}}$ & ${\sigma^2_2}$ & ${\sigma^2_{2_{err}}}$ & ${\phi_1}$ & ${\phi_2}$ \\ \hline
			5 & 0.0065 & 5.79 $\times 10^{-5}$ & -0.0032 & 5.10 $\times 10^{-5}$ & 0.0013 & 0.0015 & 0.0032 & 4.27 $\times 10^{-5}$ & 0.0124 & 8.72 $\times 10^{-5}$ & -0.0013 & 8.58 $\times 10^{-5}$ & 0.2462 & 0.0041 & 0.0029 & 6.33 $\times 10^{-5}$ & 0.57 & 0.43 \\ \hline
			
			6 & 0.0061 & 6.52 $\times 10^{-5}$ & -0.0030 & 5.55 $\times 10^{-5}$ & 0.0053 & 0.0016 & 0.0030 & 4.67 $\times 10^{-5}$ & 0.0118 & 9.39 $\times 10^{-5}$ & -0.0017 & 9.10 $\times 10^{-5}$ & 0.2072 & 0.0042 & 0.0031 & 6.83 $\times 10^{-5}$ & 0.53 & 0.47 \\ \hline
			
			7 & 0.0057 & 7.58 $\times 10^{-5}$ & -0.0028 & 6.27 $\times 10^{-5}$ & 0.0023 & 0.0018 & 0.0028 & 5.31 $\times 10^{-5}$ & 0.0110 & 1.04 $\times 10^{-4}$ & -0.0019 & 9.42 $\times 10^{-5}$ & 0.1808 & 0.0044 & 0.0031 & 7.14 $\times 10^{-5}$ & 0.48 & 0.52 \\ \hline
			
			8 & 0.0057 & 8.38 $\times 10^{-5}$ & -0.0027 & 6.64 $\times 10^{-5}$ &  0.0081 & 0.0019 & 0.0027 & 5.64 $\times 10^{-5}$ & 0.0108 & 1.29 $\times 10^{-4}$ & -0.0018 & 1.12 $\times 10^{-4}$ & 0.1784 & 0.0053 & 0.0030 & 8.57 $\times 10^{-5}$ & 0.54 & 0.46 \\ \hline
			
			9 & 0.0055 & 9.51 $\times 10^{-5}$ & -0.0027 & 7.28 $\times 10^{-5}$ & 0.0033 & 0.0020 & 0.0027 & 6.26 $\times 10^{-5}$ & 0.0104 & 1.46 $\times 10^{-4}$ & -0.0020 & 1.23 $\times 10^{-4}$ & 0.1592 & 0.0055 & 0.0031 & 9.66 $\times 10^{-5}$ & 0.54 & 0.46 \\ \hline
			
			10 & 0.0051 & 9.47 $\times 10^{-5}$ &-0.0024 & 7.20 $\times 10^{-5}$ & 0.0071 & 0.0021 & 0.0024 & 6.18 $\times 10^{-5}$ & 0.0102 & 1.62 $\times 10^{-4}$ & -0.0022 & 1.35 $\times 10^{-4}$ & 0.1436 & 0.0062 & 0.0032 & 1.05 $\times 10^{-4}$ & 0.54 & 0.46 \\ \hline
			
			11 & 0.0051 & 1.02 $\times 10^{-4}$ & -0.0024 & 7.49 $\times 10^{-5}$ & 0.0081 & 0.0022 & 0.0024 & 6.47 $\times 10^{-5}$ & 0.0102 & 1.89 $\times 10^{-4}$ & -0.0025 & 1.60 $\times 10^{-4}$ & 0.1304 & 0.0069 & 0.0034 & 1.26 $\times 10^{-4}$ & 0.61 & 0.39 \\ \hline
			
			12 & 0.0050 & 1.06 $\times 10^{-4}$ & -0.0023 & 7.35 $\times 10^{-5}$ & 0.0090 & 0.0021 & 0.0024 & 6.43 $\times 10^{-5}$ & 0.0103 & 2.15 $\times 10^{-4}$ & -0.0027 & 1.71 $\times 10^{-4}$ & 0.1238 & 0.0068 & 0.0035 & 1.39 $\times 10^{-4}$ & 0.63 & 0.37 \\ \hline
			
			15 & 0.0049 & 1.30 $\times 10^{-4}$ & -0.0024 & 8.80 $\times 10^{-5}$ & 0.0036 & 0.0022 & 0.0024 & 7.81 $\times 10^{-5}$ & 0.0090 & 2.68 $\times 10^{-4}$ & -0.0024 & 1.91 $\times 10^{-4}$ & 0.1047 & 0.0079 & 0.0031 & 1.57 $\times 10^{-5}$ & 0.73 & 0.27 \\ \hline
	\end{tabular}}
	\caption{\label{Table2} Results from applying pEMv2 for TRAP-labeled Pma1 data: covariances ($\bar{S}_{0}$ and $\bar{S}_{1}$, $\mu m^2$), diffusivities ($D$, $\mu m^2/s$), localization errors ($\sigma^2$, $\mu m^2$), and volume
		fractions (${\phi}$) for the two states found by pEMv2 for track lengths of $N=5$, 6, 7, 8, 9, 10, 11, 12, and 15. }
\end{table*}

Table \ref{Table2} reports the mean covariances, $\bar{S}_0$ and $\bar{S}_1$,
the mean diffusivity, $\bar{D}$ and the mean localization noise for these two diffusive states,
as well as
the fraction of the population corresponding to state 1, $\phi_1$,
the fraction of the population corresponding to state 2, $\phi_2$.
Evidently, pEMv2 indicates the existence of a relatively low diffusivity state (state 1)
with $D_1 \simeq 0.003~\mu$m$^2$s$^{-1}$ and a relatively high diffusivity state (state 2) with $D_2 \simeq 0.16~\mu$m$^2$s$^{-1}$, for a track length of 9 steps.
Both states have roughly equal representation in the overall population
with $\phi_1 \simeq 0.54$ and $\phi_2 \simeq 0.46$.

Table \ref{Table2} reveals that the diffusivity
of state 2, returned by pEMv2, decreases as the track length increases.
We hypothesize that this circumstance arises because for increasing track length, high diffusivity tracks have a
progressively higher
probability to diffuse out of the illuminated volume than lower diffusivity tracks.
Thus, a population of longer tracks is preferentially depleted of high-diffusivity tracks, pulling down the
population's mean diffusivity, compared to a population of shorter tracks.

\subsection{Mean-square displacements (MSDs) versus delay time of pEMv2-sorted trajectories}
\label{Sec:MSD}
\begin{figure} 
	\includegraphics[scale=0.455]{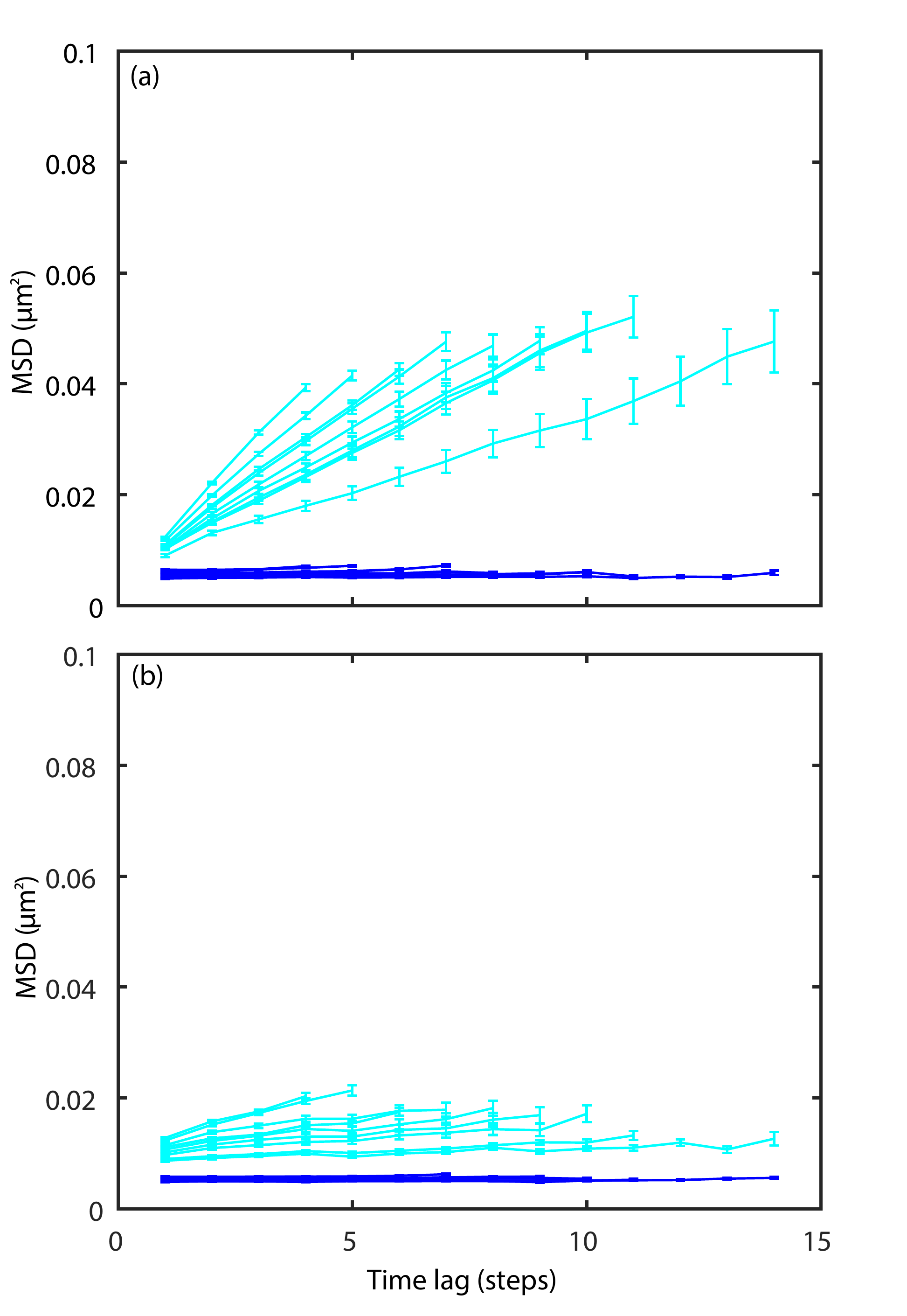}% Here is how to import EPS art
	\caption{\label{fig:MSDs} Comparison of the mean squared displacements (MSDs) for state 1 (blue) and state 2 (cyan) across varying track lengths, for experimental TRAP-labeled Pma1 (a), and experimental direct fusion data (b). In each case, the state 2 MSDs have higher slopes than state 1.  }
\end{figure}

In Sec.~\ref{Sec:pEM}, we used the pEMv2-determined mean covariances to estimate the diffusivity of each diffusive state.
Ref.~\cite{Vestergaard2014} showed that such covariance-based estimators are statistically optimal.
However, another widely-used means of
characterizing diffusivity is via the slope of the mean-square displacement (MSD) versus time delay,
which is a linear function for simple diffusion.
Fig.~\ref{fig:MSDs}(a) plots the mean-square MSDs for TRAP-labeled Pma1 trajectories
for each state versus the number of steps.
The MSDs for tracks corresponding to state 1 are shown in blue, while
the MSDs for tracks corresponding to state 2 are shown in cyan.
Evidently, the MSDs for state 1 are conspicuously  constant versus time.
This observation emphasizes that
state 1 really is essentially immobile.
By contrast, it is clear that the 
state-2 MSDs increase approximately linearly versus time,
consistent with what is expected for simple diffusion.
It is also apparent that the slopes of the state-2 MSDs tend to decrease with increasing track length,
corresponding to an apparently decreasing diffusivity with increasing track length,
mirroring  the behavior observed for the covariance-based estimates of the diffusivity.

\begin{figure} [t!]
	\includegraphics[width=3.5in]{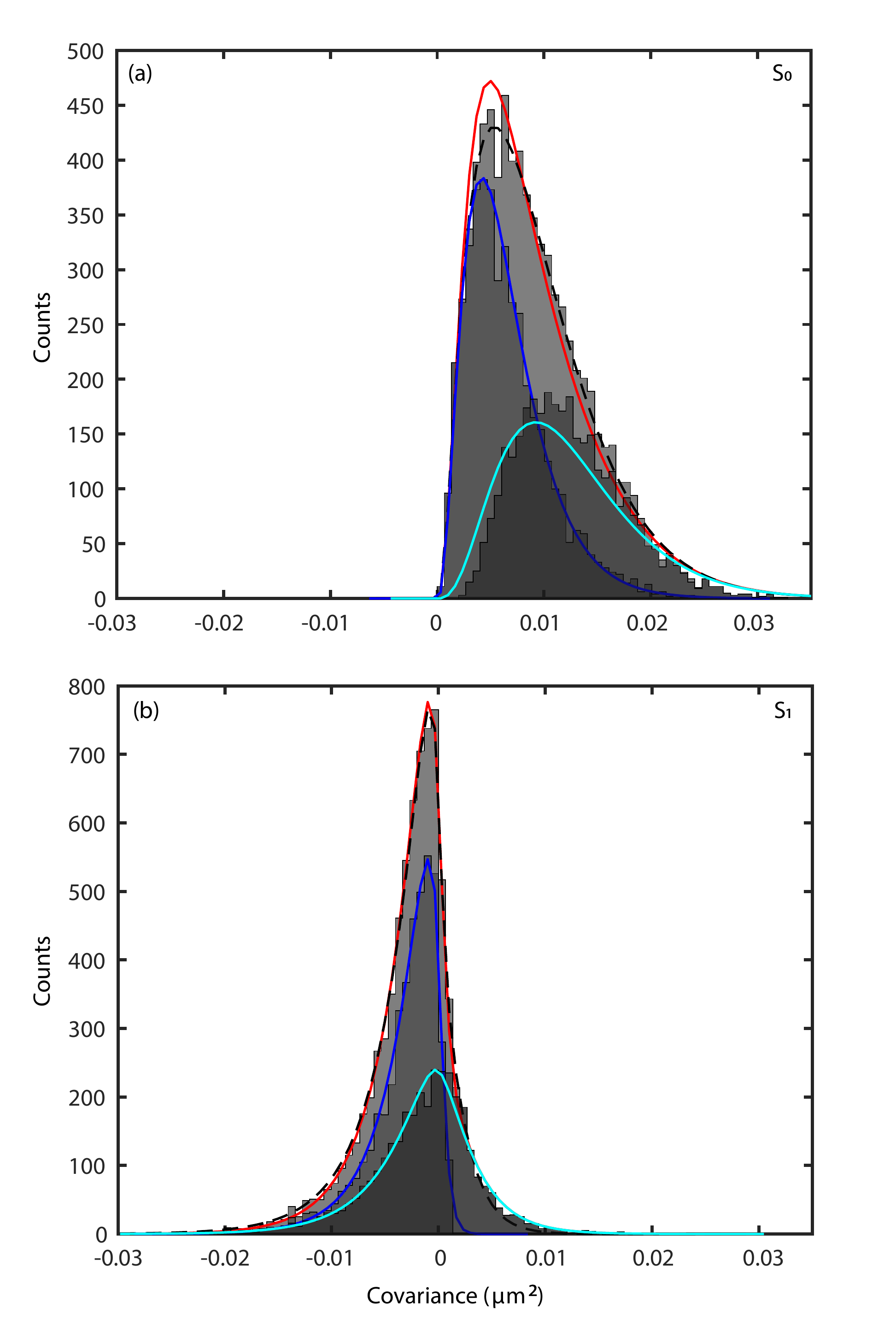}
	\caption{\label{SplitLength5} Distribution of covariance elements (a) $S_0$ and (b) $S_1$, for
		experimental  5-step  TRAP-labeled Pma1 tracks. The unsorted covariance distribution (light grey histogram) is plotted with the theoretical two-component curve (red curve) given by Eq.~\ref{EQ-11}, and the fitted two-component curve (black dashed line). The tracks are sorted into two distributions representing the two distinct diffusive states found by pEMv2 (dark grey and darker grey histograms), and plotted with their single theory curves (blue and cyan curves), given by Eq.~\ref{EQ-16}.
	}  
\end{figure}

\begin{figure} [t!]
	\includegraphics[width=3.5in]{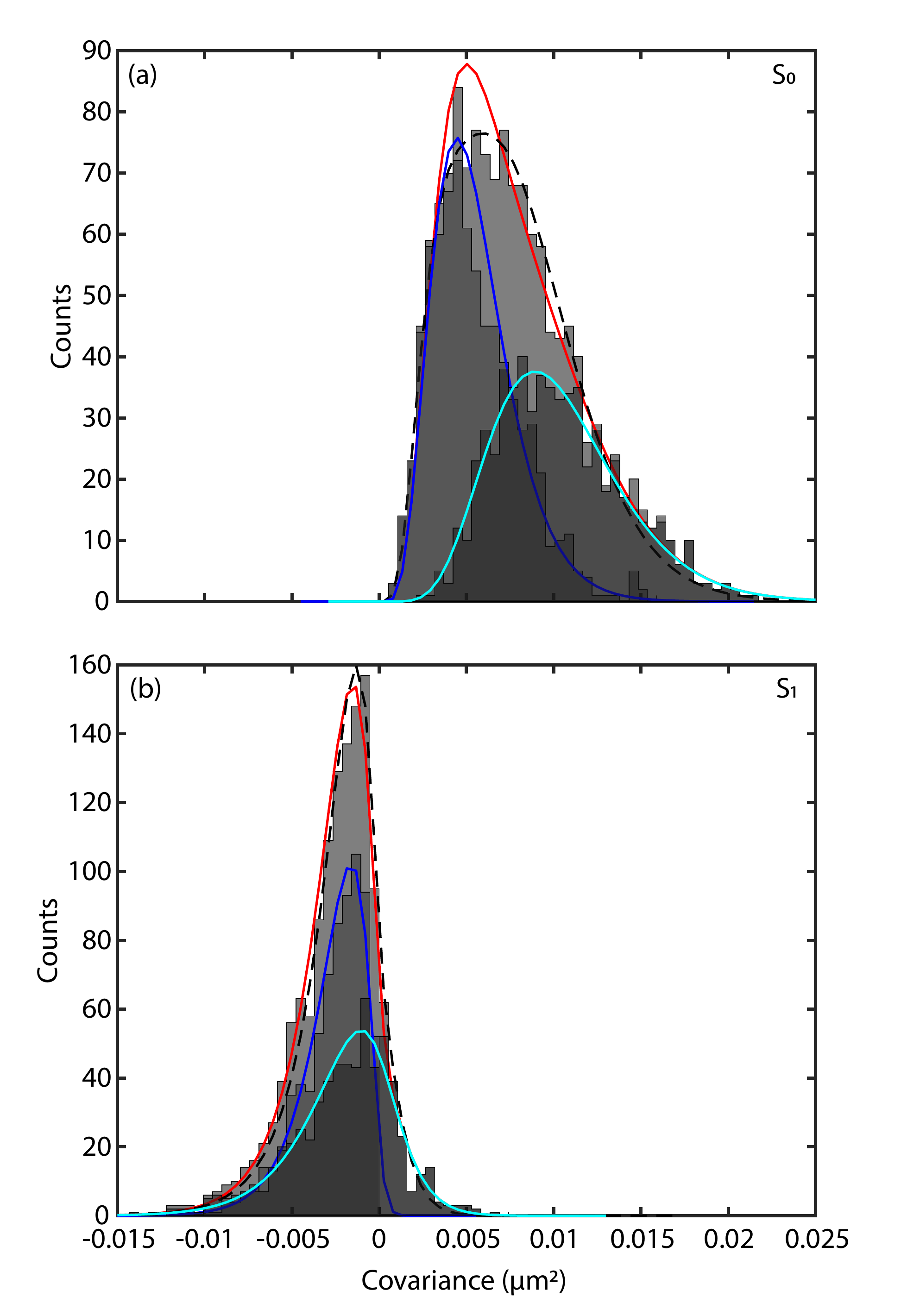}
	\caption{\label{fig:ftirf_cov} Distribution of covariance elements (a) $S_0$ and (b) $S_1$, for
		experimental  9-step  TRAP-labeled Pma1 tracks. The unsorted covariance distribution (light grey histogram) is plotted with the theoretical two-component curve (red curve) given by Eq.~\ref{EQ-11}, and the fitted two-component curve (black dashed line). The tracks are sorted into two distributions representing the two distinct diffusive states found by pEMv2 (dark grey and darker grey histograms), and plotted with their single theory curves (blue and cyan curves), given by Eq.~\ref{EQ-16}.
	}  
\end{figure}

\begin{figure} [t!]
	\includegraphics[width=3.5in]{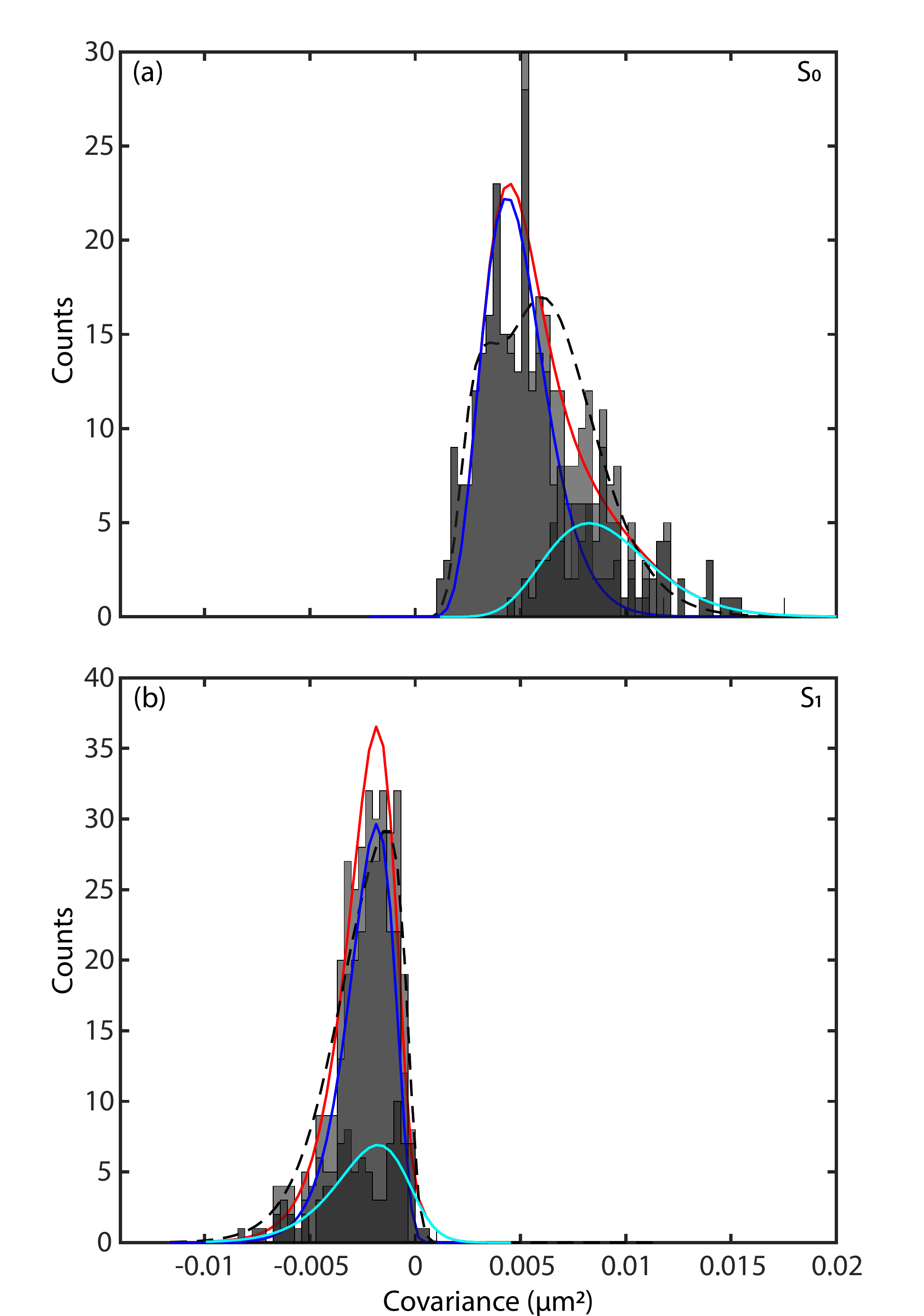}
	\caption{\label{SplitLength15} Distribution of covariance elements (a) $S_0$ and (b) $S_1$, for
		experimental  15-step  TRAP-labeled Pma1 tracks. The unsorted covariance distribution (light grey histogram) is plotted with the theoretical two-component curve (red curve) given by Eq.~\ref{EQ-11}, and the fitted two-component curve (black dashed line). The tracks are sorted into two distributions representing the two distinct diffusive states found by pEMv2 (dark grey and darker grey histograms), and plotted with their single theory curves (blue and cyan curves), given by Eq.~\ref{EQ-16}.
	}  
\end{figure}

\subsection{Covariance distributions of pEMv2-sorted TRAP-labeled Pma1 trajectories agree with theory}
\label{Sec:experimental_ftirf}
To further assess the performance of pEMv2, we first sought to compare the theoretical covariance distributions
for each state, conditioned on their respective experimental mean covariances, $\bar{S}_0$ and $\bar{S}_1$,
to the corresponding experimental distributions.
Such comparisons are presented in
Fig.~\ref{SplitLength5},  Fig.~~\ref{fig:ftirf_cov}, and Fig.~\ref{SplitLength15}, which plot both experimental and theoretical covariance distributions
%, $S_0$ (Fig.~\ref{fig:ftirf_cov}(a)) and $S_1$ (Fig.~\ref{fig:ftirf_cov}(b)), 
for 5-step, 9-step, and 15-step trajectories, respectively,
sorted into states 1 and 2 and for the overall population of unsorted trajectories.
Stepped histograms represent the experimental covariance distributions.
Smooth lines are the theoretical covariance distributions with
blue and cyan corresponding to the distributions (Eq.~\ref{EQ-16}) for  state 1 and state 2, respectively,
and red corresponding to the two-component distribution (Eq.~\ref{EQ-11}) for the overall, unsorted
population.  In all cases,the theory and experiment appear to be in good agreement, despite the fact that no fitting is involved: The only input for the theoretical curves in these figures are the experimental mean values
of  $\bar{S}_0$ and $\bar{S}_1$.
The fact that the distributions to emerge from the pEMv2-sorted populations agree well with
what is expected theoretically, without any fitting, strongly supports that pEMv2 is correctly characterizing the diffusive
behavior of TRAP-labeled Pma1. Quantitative comparisons of the pEMv2-sorted experimental distributions and theory using the Kolmogorov-Smirnov test are described in Supplementary Material 1A.

\begin{figure} 
	\includegraphics[width=3.5in]{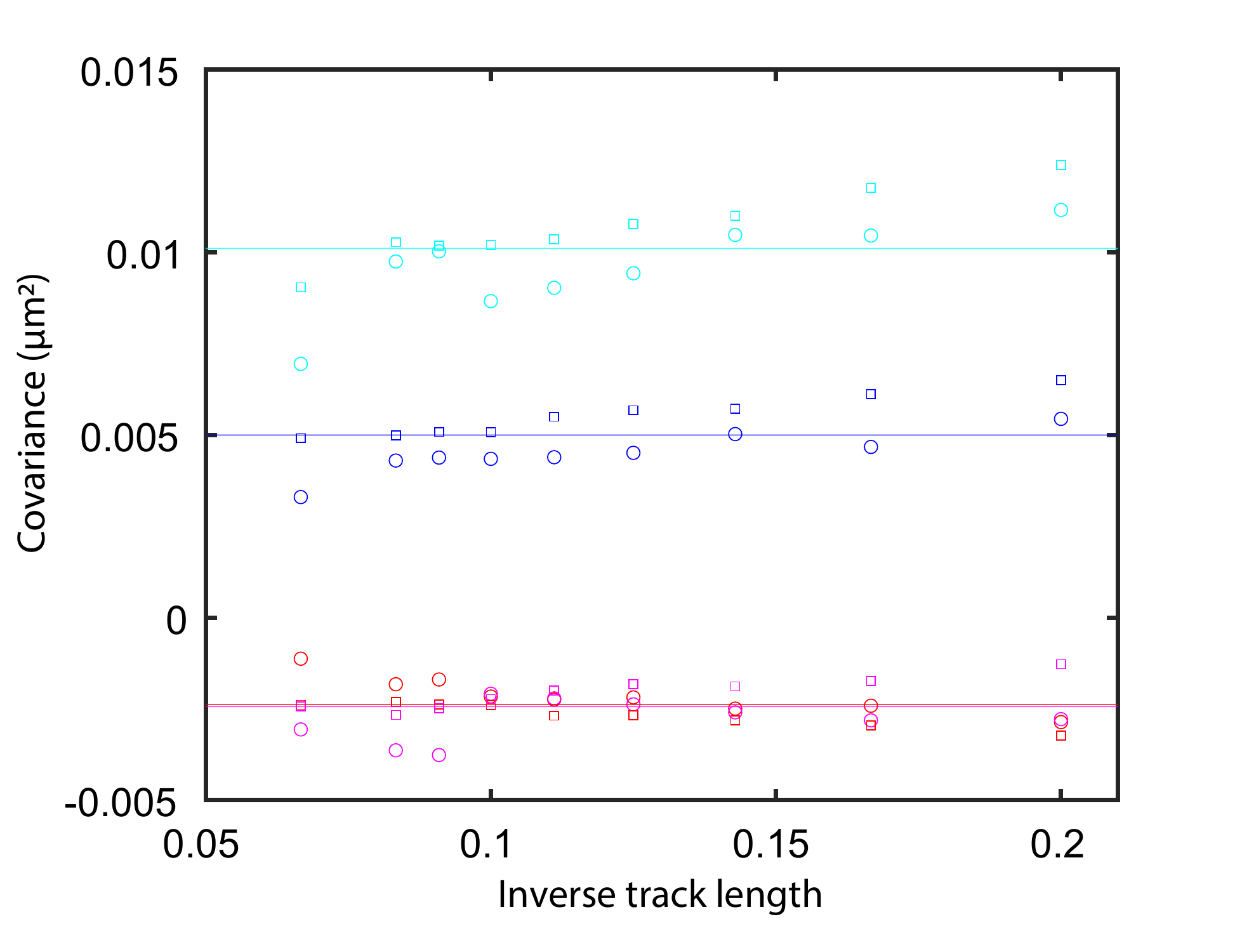}
	\caption{\label{fig:s01_vs_tracklength} Plot comparing the fitted $S_0$ and $S_1$ values (circles) to the pEMv2 found values (squares), versus inverse track length, for experimental TRAP-labeled Pma1. The $S_0$ values are in blue and cyan for states 1 and 2, respectively. The $S_1$ values are in red and magenta for states 1 and 2, respectively. Solid lines represent guides to the data, and are calculated by taking the average of the $S$ values for each state over the track lengths. }
\end{figure}

\subsection{pEMv2 finds two diffusive states for Pma1-mEos3.2 direct-fusion trajectories}
\label{sec:direct fusion}
\begin{figure} %BIC N=9 FTIRF
	\includegraphics[scale=0.455]{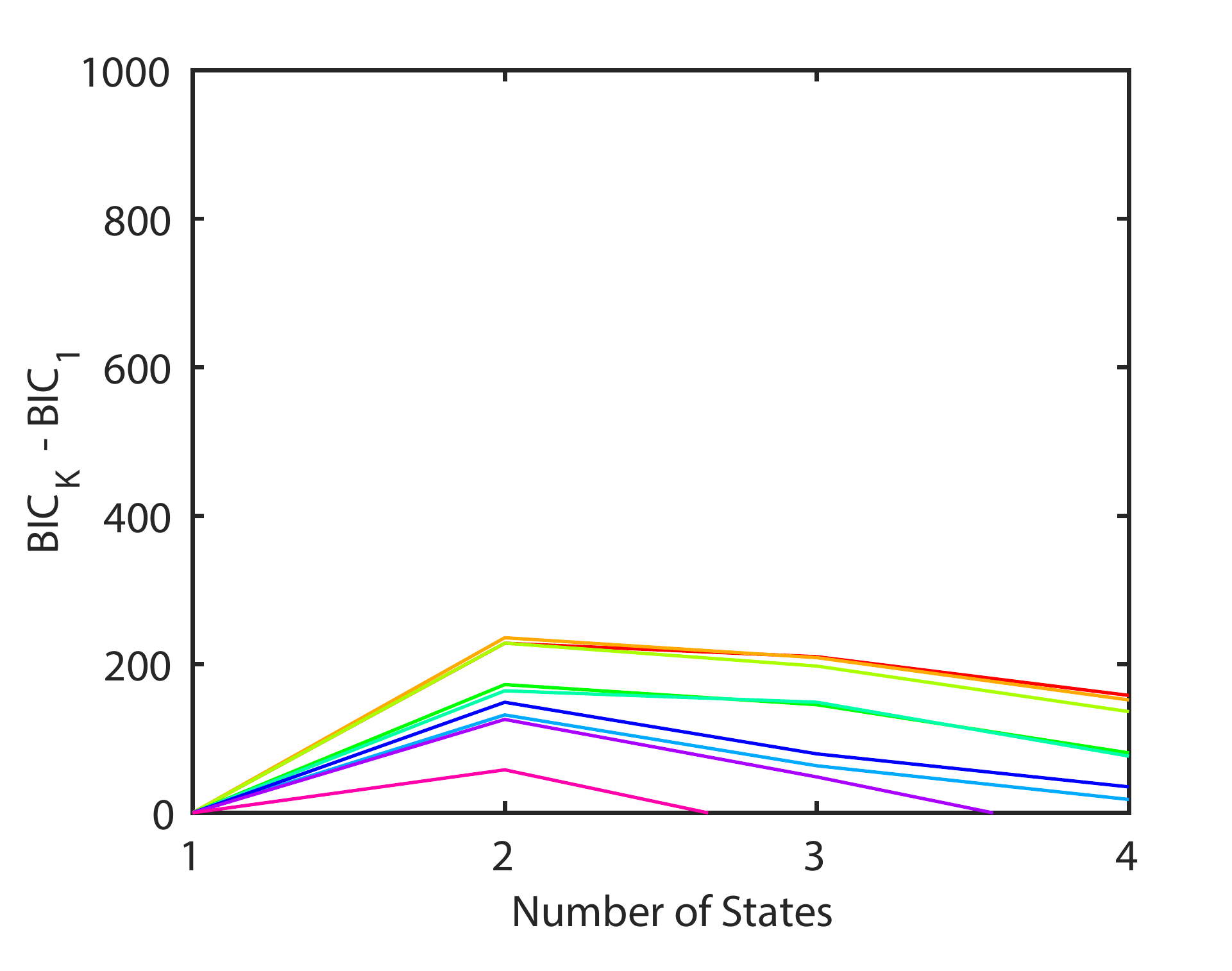}% Here is how to import EPS art
	\caption{\label{fig:dtirf_bic}
		Relative Bayesian Information Criterion (BIC$_K$-BIC$_1$) values
		versus number of diffusive states for Pma1-mEos3.2 direct fusion trajectories of
		length 5 (red), 6 (orange), 7 (lime green), 8 (green), 9, (blue-green), 10 (blue), 11 (dark blue), 12 (purple), and 15 (magenta).
		\color{black} 
		The highest relative BIC value for each track length occurs for two states, irrespective of track length. }
\end{figure}

Next, we  applied pEMv2 and the theory of Sec.~\ref{Sec:Theory} to Pma1-mEos3.2 direct-fusion trajectories.
As shown in Fig.~\ref{fig:dtirf_bic},
for track lengths of 5, 6, 7, 8, 9, 10, 11, 12, and 15, the BIC was maximized for two states.
However, the increase in the BIC, going from one state to two states, is much less in this case ($BIC_2-BIC_1=250$), than for
TRAP-labeled Pma1 trajectories ($BIC_2-BIC_1=1000$, Fig.~\ref{fig:ftirf_bic}), indicating that pEMv2's preference for a two state
description over a one-state description is much less for
Pma1-mEos3.2 than for TRAP-labeled Pma1. 
%Fig.~\ref{ExampleTracks_dtirf} shows eight example 12-step tracks for each of Pma1-mEos3.2's two states, with state 1 shown in blue, and state 2 in cyan.

\begin{table*}
	%\begin{center}
	\centering\resizebox{\textwidth}{!}
	{\begin{tabular}{| c | c | c | c | c | c | c | c | c | c | c | c | c | c | c | c | c | c | c |}
			\hline
			{Track Length} & ${\bar{S}_{01}}$ & ${\bar{S}_{01_{err}}}$ & ${\bar{S}_{11}}$ & ${\bar{S}_{02_{err}}}$ & ${D_1}$ & ${D_{1_{err}}}$ & ${\sigma^2_1}$ & ${\sigma^2_{1_{err}}}$ & ${\bar{S}_{02}}$  & ${\bar{S}_{02_{err}}}$ & ${\bar{S}_{12}}$  & ${\bar{S}_{12_{err}}}$ & ${D_2}$ & ${D_{2_{err}}}$ & ${\sigma^2_2}$ & ${\sigma^2_{2_{err}}}$ & ${\phi_1}$ & ${\phi_2}$ \\ \hline
			5 & 0.0056 & 4.65 $\times 10^{-5}$ & -0.0029 & 4.19 $\times 10^{-5}$ & -0.0021 & 0.0013 & 0.0028 & 3.47 $\times 10^{-5}$ & 0.0128 & 1.64 $\times 10^{-4}$ & -0.0048 & 1.78 $\times 10^{-4}$ & 0.081 & 0.0063 & 0.0053 & 1.41 $\times 10^{-4}$ & 0.84 & 0.16 \\ \hline
			
			6 & 0.0058 & 5.17 $\times 10^{-5}$ & -0.0029 & 4.27 $\times 10^{-5}$ & 0.0012 & 0.0012 & 0.0029 & 3.61 $\times 10^{-5}$ & 0.0123 & 2.00 $\times 10^{-4}$ & -0.0046 & 2.08 $\times 10^{-4}$ & 0.079 & 0.0071 & 0.0051 & 1.673 $\times 10^{-4}$ & 0.89 & 0.11 \\ \hline
			
			7 & 0.0053 & 5.03 $\times 10^{-5}$ & -0.0027 & 4.26 $\times 10^{-5}$ & 0.00075 & 0.0013 & 0.0027 & 3.57 $\times 10^{-5}$ & 0.0111 & 1.57 $\times 10^{-4}$ & -0.0047 & 1.56 $\times 10^{-4}$ & 0.0448 & 0.0053 & 0.0050 & 1.26 $\times 10^{-4}$ & 0.82 & 0.18 \\ \hline
			
			8 & 0.0057 & 5.76 $\times 10^{-5}$ & -0.0028 & 4.56 $\times 10^{-5}$ &  0.0021 & 0.0013 & 0.0028 & 3.89 $\times 10^{-5}$ & 0.0115 & 2.40 $\times 10^{-4}$ & -0.0045 & 2.32 $\times 10^{-4}$ & 0.061 & 0.0075 & 0.0049 & 1.90 $\times 10^{-4}$ & 0.91 & 0.09 \\ \hline
			
			9 & 0.0055 & 5.98 $\times 10^{-5}$ & -0.0027 & 4.72 $\times 10^{-5}$ & 0.0015 & 0.0013 & 0.0027 & 4.03 $\times 10^{-5}$ & 0.0107 & 2.14 $\times 10^{-4}$ & -0.0046 & 2.10 $\times 10^{-4}$ & 0.040 & 0.0068 & 0.0048 & 1.70 $\times 10^{-4}$ & 0.89 & 0.11 \\ \hline
			
			10 & 0.0054 & 6.18 $\times 10^{-5}$ &-0.0026 & 4.82 $\times 10^{-5}$ & 0.0032 & 0.0014 & 0.0027 & 4.13 $\times 10^{-5}$ & 0.0102 & 2.37 $\times 10^{-4}$ & -0.0044 & 2.16 $\times 10^{-4}$ & 0.035 & 0.0066 & 0.0046 & 1.79 $\times 10^{-4}$ & 0.89 & 0.11 \\ \hline
			
			11 & 0.0053 & 6.30 $\times 10^{-5}$ & -0.0026 & 4.81 $\times 10^{-5}$ & 0.0020 & 0.0013 & 0.0026 & 4.14 $\times 10^{-5}$ & 0.0100 & 2.32 $\times 10^{-4}$ & -0.0042 & 2.05 $\times 10^{-4}$ & 0.032 & 0.0064 & 0.0044 & 1.70 $\times 10^{-4}$ & 0.86 & 0.14 \\ \hline
			
			12 & 0.0049 & 5.90 $\times 10^{-5}$ & -0.0024 & 4.41 $\times 10^{-5}$ & 0.0036 & 0.0013 & 0.0024 & 3.80 $\times 10^{-5}$ & 0.0090 & 1.64 $\times 10^{-4}$ & -0.0042 & 1.45 $\times 10^{-4}$ & 0.0136 & 0.0045 & 0.0043 & 1.20 $\times 10^{-4}$ & 0.77 & 0.23 \\ \hline
			
			15 & 0.0050 & 6.80 $\times 10^{-4}$ & -0.0025 & 5.16 $\times 10^{-5}$ & 0.0012 & 0.0014 & 0.0025 & 4.45 $\times 10^{-5}$ & 0.0087 & 2.01 $\times 10^{-4}$ & -0.0040 & 1.80 $\times 10^{-4}$ & 0.0166 & 0.00549 & 0.0041 & 1.50 $\times 10^{-4}$ & 0.82 & 0.18 \\ \hline
	\end{tabular}}
	\caption{\label{Tabledirectfusion} Results from applying pEMv2 for direct fusion data: covariances ($\bar{S}_{0}$ and $\bar{S}_{1}$, $\mu m^2$), diffusivities ($D$, $\mu m^2/s$), localization errors ($\sigma^2$, $\mu m^2$), and volume
		fractions (${\phi}$) for the two states found by pEMv2 for track lengths of $N=5$, 6, 7, 8, 9, 10, 11, 12, and 15. }
\end{table*}

Table \ref{Tabledirectfusion} reports the mean covariances, $\bar{S}_0$ and $\bar{S}_1$,
the mean diffusivity, $\bar{D}$ and the mean localization noise for the two diffusive states, found in this case,
as well as
the fraction of the population corresponding to state 1, $\phi_1$,
the fraction of the population corresponding to state 2, $\phi_2$.
From this table, it is clear that
pEMv2 finds that Pma1-mEos3.2 is overall significantly less mobile than TRAP-labeled Pma1.
First,  the immobile state (state 1)
%with $D_1 \simeq 0.003~\mu$m$^2$s$^{-1}$
constitutes about 90\% of the total population.
Second,  the diffusivity of the higher diffusivity state (state 2) is just $D_2 \simeq 0.04~\mu$m$^2$s$^{-1}$,
several times smaller than the diffusivity of TRAP-labeled Pma1's higher-diffusivity state,
for which
$D_2 \simeq 0.15~\mu$m$^2$s$^{-1}$.
Hence, the more-mobile second state in this labeling method, is still far less mobile than the mobile state for the indirect labeling.

\begin{figure} 
	\includegraphics[scale=0.450]{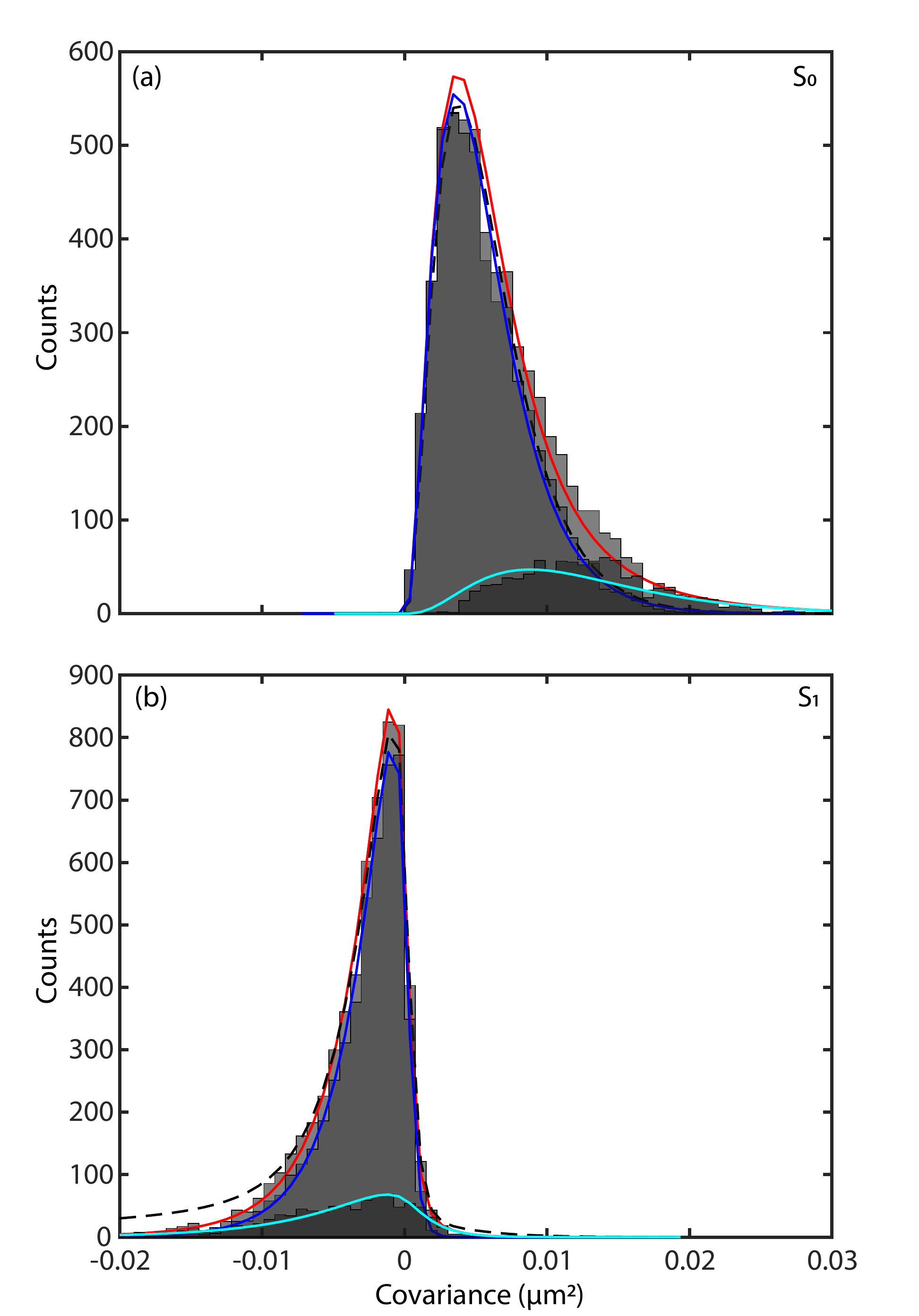}
	\caption{\label{SplitLength5DirectFusion} Distribution of covariance elements, $S_0$ (a) and $S_1$ (b), for experimental direct fusion tracks of length 5 steps. The unsorted covariance distribution (light grey histogram) is plotted with the theoretical two-component curve (red curve) given by Eq.~\ref{EQ-11}, and the fitted two-component curve (black dashed line). The tracks are sorted into two distributions representing the two distinct diffusive states found by pEMv2 (dark grey and darker grey histograms), and plotted with their single theory curves (blue and cyan curves), given by Eq.~\ref{EQ-16}.
	}  
\end{figure}

\begin{figure} 
	\includegraphics[scale=0.450]{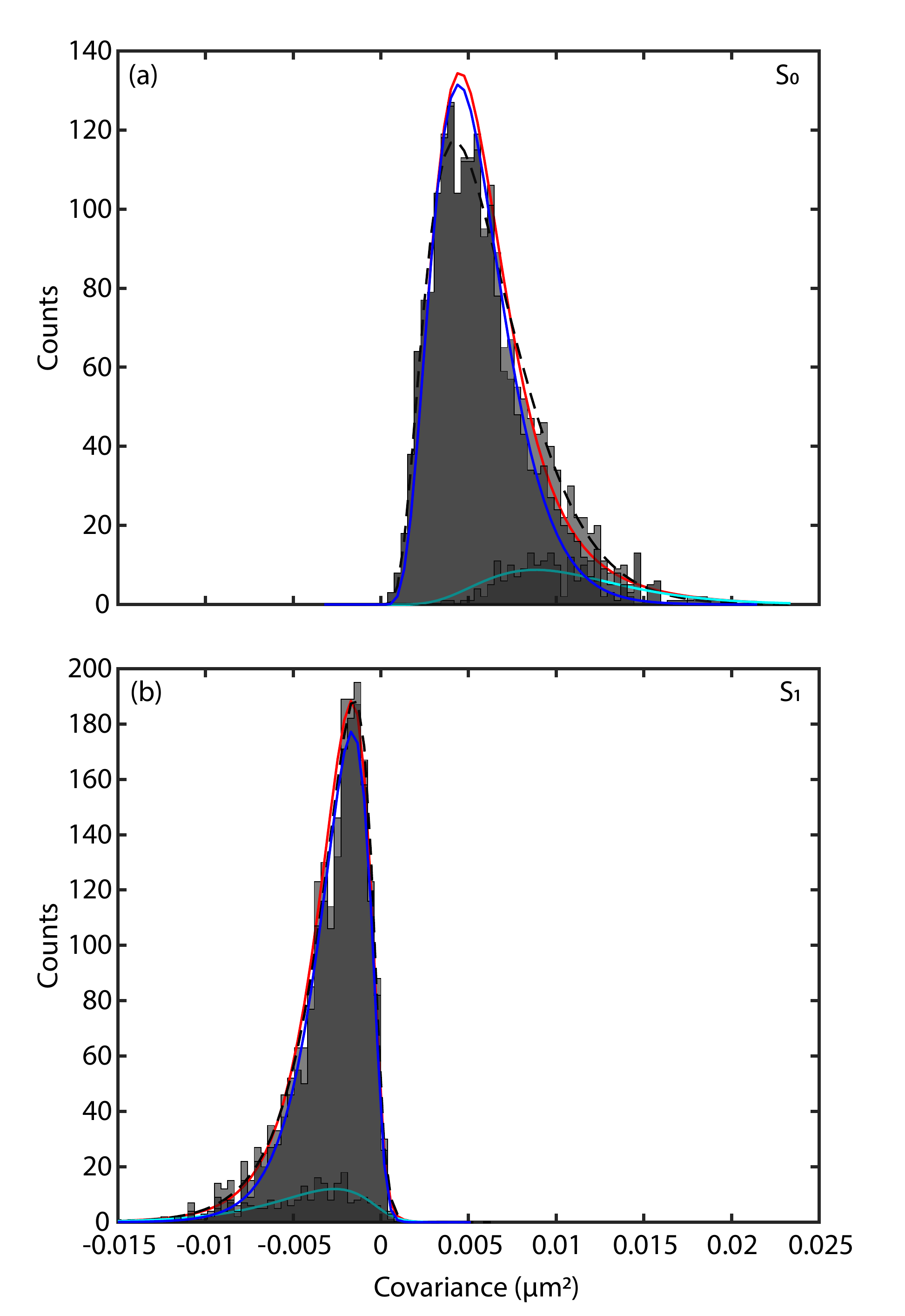}
	\caption{\label{fig:dtirf_cov} Distribution of covariance elements, $S_0$ (a) and $S_1$ (b), for experimental direct fusion tracks of length 9 steps. The unsorted covariance distribution (light grey histogram) is plotted with the theoretical two-component curve (red curve) given by Eq.~\ref{EQ-11}, and the fitted two-component curve (black dashed line). The tracks are sorted into two distributions representing the two distinct diffusive states found by pEMv2 (dark grey and darker grey histograms), and plotted with their single theory curves (blue and cyan curves), given by Eq.~\ref{EQ-16}.
	}  
\end{figure}

\begin{figure} 
	\includegraphics[scale=0.450]{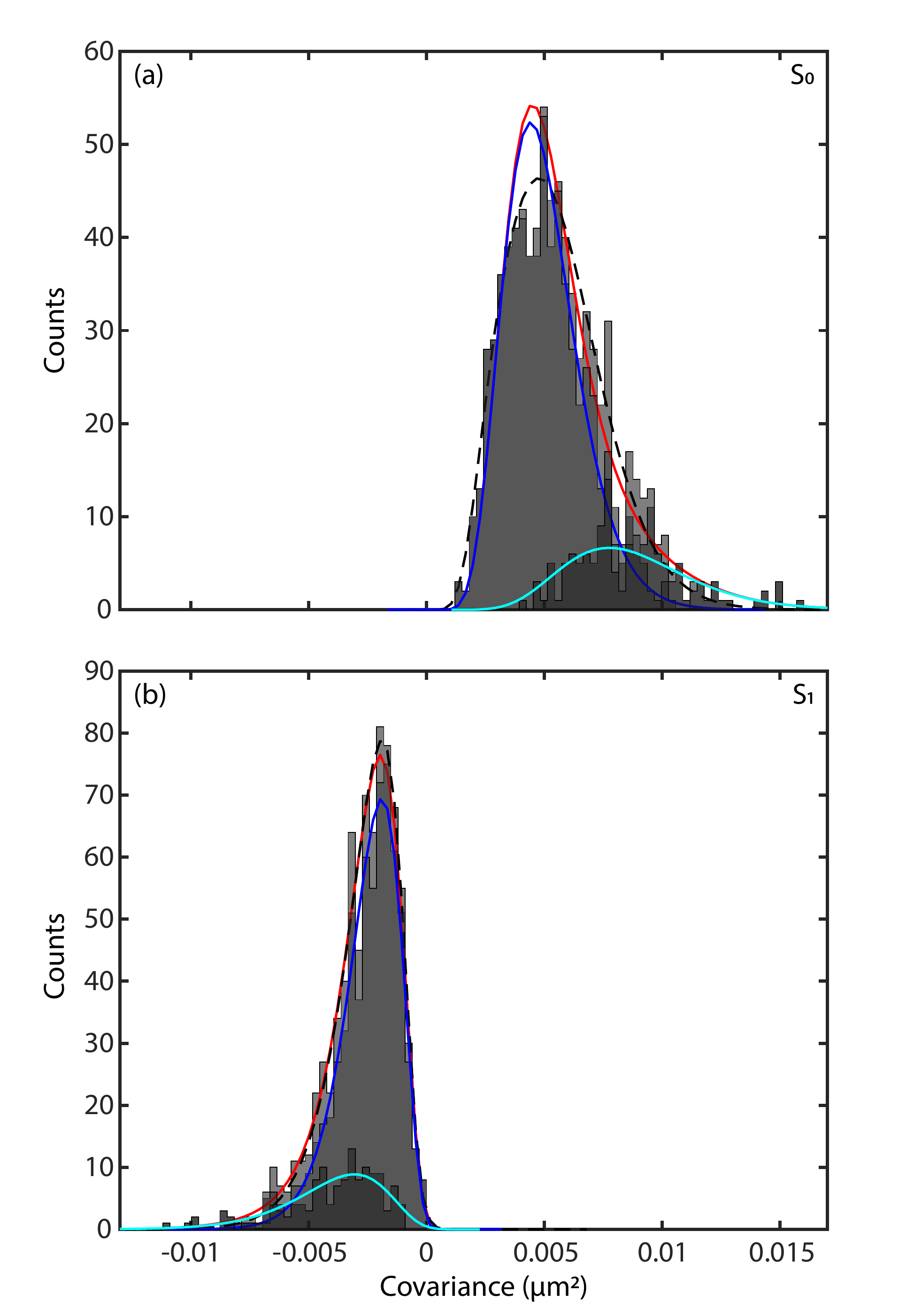}
	\caption{\label{SplitLength15DirectFusion} Distribution of covariance elements, $S_0$ (a) and $S_1$ (b), for experimental direct fusion tracks of length 15 steps. The unsorted covariance distribution (light grey histogram) is plotted with the theoretical two-component curve (red curve) given by Eq.~\ref{EQ-11}, and the fitted two-component curve (black dashed line). The tracks are sorted into two distributions representing the two distinct diffusive states found by pEMv2 (dark grey and darker grey histograms), and plotted with their single theory curves (blue and cyan curves), given by Eq.~\ref{EQ-16}.
	}  
\end{figure}

%Fig.~\ref{fig:dtirf_cov} shows the distribution of covariances for Pma1-mEos3.2 trajectories for a track length of 9 steps. The corresponding CDFs are shown in Fig.~\ref{fig:dtirf_cdf_states_9} and Fig.~\ref{fig:dtirf_cdf_fit}.  For the direct fusion case too, we find good agreement between theory and experiment both for the sorted and unsorted distributions. 

Fig.~\ref{SplitLength5DirectFusion},  Fig.~\ref{fig:dtirf_cov} , and Fig.~\ref{SplitLength15DirectFusion}
compare the  experimental covariance distributions for Pma1-mEos3.2 trajectories with track lengths of 5, 9, and 15 steps,
respectively, to the corresponding theoretical predictions.
In each figure, there appears to be good agreement between theory and data, even though no fitting
is involved.
The fact that the distributions to emerge from the pEMv2-sorted populations agree well with
what is expected theoretically supports that pEMv2 is correctly characterizing the diffusive
behavior of Pma1-mEOs3.2. Quantitative comparisons of the pEMv2-sorted experimental distributions and theory, using the Kolmogorov-Smirnov test, are described in Supplementary Material 1B.

Fig.~\ref{fig:D_exp}(b) shows the diffusivity distributions for Pma1-mEOs3.2 trajectories.
Two features stand out from this figure in comparison to Fig.~\ref{fig:D_exp}(a).
First, the population fraction of the higher diffusivity state (state 2) is significantly less for the direct fusion than
for TRAP-labeled Pma1.
Second, the mean diffusivity of the higher diffusivity state (state 2) is significantly less for the direct fusion
($\sim 0.05~\mu$m$^2$s$^{-1}$)than
for TRAP-labeled Pma1 ($\sim 0.16~\mu$m$^2$s$^{-1}$).
These two features make the overall diffusivity of Pma1-mEos3.2
significantly smaller than that of TRAP-labeled Pma1.

\section{Conclusion}
\label{conclusion}
Two primary conclusions are to be drawn from this work.
First,  the diffusive behavior of Pma1
can be described convincingly in terms of two discrete diffusive states, each with its own diffusive properties, sorted by the pEMv2 software, and validated using our theoretical covariance distributions. 
Specifically, we investigated two differently labeled versions of Pma1 at the single molecule level using single particle tracking
under TIRF illumination.
Using a machine-learning-based approach that identifies and sorts a population of trajectories into a discrete
number of diffusive states, pEMv2, we found that a minimally-modified version of Pma1,
with a C-terminal five-amino-acid tag
that reversibly binds to a TRAP-mEos3.2 fusion, shows comparable population fractions of
a mobile state, corresponding to simple diffusion with a diffusion constant consistent with what is expected for a membrane protein, and
an essentially immobile state.
By contrast, we found that a Pma1-mEos3.2 direct fusion is overwhelmingly in an immobile state, and the small population fraction, assigned to a more mobile state by pEMv2, has a diffusion coefficient several times smaller than the diffusion coefficient
of mobile TRAP-labeled Pma1.
A comparison between the experimental pEMv2-sorted covariance and diffusivity distributions, conditioned on the
experimental mean covariances of the sorted tracks, 
and the corresponding theoretical distributions that we derived, shows
overall good agreement without any fitting, providing strong support for pEMv2-based sorting.
Additionally, we note that the method used to label a protein for microscopy visualization in living cells can
affect the protein's diffusive behavior in a substantial fashion
with unknown consequences for its biological function.
Further studies are required to determine which labeling method more closely represents the motion of the unlabeled protein. 
Second, we found that both variants of Pma1 studied in this work, TRAP-labeled Pma1 and Pma1-mEos3.2, are monomers {\em in vivo}. This result is noteworthy because in vitro preparations for electron microscopy show Pma1 forms hexamers. We estimated the lifetime of the TRAP-peptide bound-state to be 0.15~s.
An open question remains as to whether Pma1's two diffusive states are significant for Pma1's biological function, and if so
what their roles might be.

\bmhead{Acknowledgments}
	M.L.P.B. and S.E.P. contributed equally to this work.
	M.L.P.B. conceived the project, analyzed data, and wrote the paper;
	S.E.P. conceived the project, collected data, analyzed data, and wrote the paper;
	Y.Z. conceived the project, collected data, and wrote the paper;
	M.H. conceived the project, constructed the yeast strains, and wrote the paper;
	J.B. conceived the project and wrote the paper;
	L.R.  conceived the project  and wrote the paper;
	S.G.J.M. conceived the project and wrote the paper.
	This research was supported by 
	the National Institutes of Health via NIH R01 GM118528 and NIH P30 DK045735.
	M.L.P.B. was supported by NIH T32EB019941 and the NSF GRFP.
	S.E.P. was supported by the NSF GRFP and NSF PHYS 1522467.
	J. B. discloses significant financial interest in Bruker Corp., Hamamatsu Photonics, and Panluminate Inc.

\bibliography{apssamp_MLB_SM042220}

\end{document}